\documentclass[fleqn,usenatbib]{mnras}

\usepackage{newtxtext,newtxmath}
\usepackage[T1]{fontenc}
\usepackage{ae,aecompl}
\usepackage[normalem]{ulem}

\usepackage{graphicx}	% Including figure files
\usepackage{amsmath}	% Advanced maths commands
\usepackage{amssymb}	% Extra maths symbols
\usepackage{etoolbox}
\makeatletter
\patchcmd\@combinedblfloats{\box\@outputbox}{\unvbox\@outputbox}{}{%
	\errmessage{\noexpand\@combinedblfloats could not be patched}%
}%
\makeatother
\usepackage{color} % To colour text
\usepackage[normalem]{ulem} % To strike out text
\usepackage{hyperref} 
\usepackage{tikz}
\usetikzlibrary{shapes.geometric, arrows}

\tikzstyle{startstop} = [rectangle, rounded corners, minimum width=2cm, minimum height=1cm,text centered, draw=black, text width=2.0cm, fill=red!30]
\tikzstyle{io} = [trapezium, trapezium left angle=70, trapezium right angle=110, minimum width=2cm, minimum height=1cm, text width=2.0cm, text centered, draw=black, fill=blue!30]
\tikzstyle{process} = [rectangle, minimum width=2cm, minimum height=1cm, text centered, text width=2.1cm, draw=black, fill=orange!30]
\tikzstyle{decision} = [diamond, minimum width=2.5cm, minimum height=1cm, text centered, draw=black, fill=green!30]
\tikzstyle{arrow} = [ultra thick,->,>=stealth]

\newcommand{\AHF}{\mbox{AHF}}

\newcommand{\FullName}{\mbox{MErger Graph Algorithm}}
\newcommand{\Name}{\mbox{MEGA}}

\newcommand{\halos}{\mbox{haloes}}
\newcommand{\Halos}{\mbox{Haloes}}
\newcommand{\Graph}{\mbox{\Name-Graph}}
\newcommand{\Tree}{\mbox{\Name-Tree}}

\newcommand{\Np}{\mbox{$N_\mathrm{P}$}}
\newcommand{\Rhalo}{\mbox{$\sigma_\mathrm{r}$}}
\newcommand{\subfind}{\mbox{\sc Subfind}}
\newcommand{\subhalo}{\mbox{subhalo}}
\newcommand{\subhalos}{\mbox{subhaloes}}
\newcommand{\Subhalo}{\mbox{Subhalo}}
\newcommand{\Subhalos}{\mbox{Subhaloes}}

\newenvironment{shortitem}
{\begin{list}{$\bullet$}{\topsep=0pt\itemsep=0pt\parsep=0pt\parskip=0pt\leftmargin=12pt}}
{\end{list}}

\newcommand{\App}[1]{Appendix~\ref{sec:#1}}
\newcommand{\Eq}[1]{Equation~\ref{eq:#1}}
\newcommand{\Fig}[1]{Figure~\ref{fig:#1}}
\newcommand{\Sec}[1]{Section~\ref{sec:#1}}

\defcitealias{Srisawat13}{SMT13}

\title[Merger graphs]{MEGA: Merger graphs of structure formation}

\author[Roper et al.]
{William J. Roper$^{1}$\thanks{E-mail: w.roper@sussex.ac.uk},
Peter A. Thomas$^{1}$,
Chaichalit Srisawat$^2$
\\
$^{1}$Astronomy Centre, University of Sussex, Falmer, Brighton BN1 9QH, UK\\
$^2$Center for Astrophysics and Cosmology, Science Institute, University of Iceland, Dunhagi 5, 107 Reykjavik, Iceland\\
}

\date{Accepted XXX. Received YYY; in original form ZZZ}

\pubyear{2020}

\begin{document}
\label{firstpage}
\pagerange{\pageref{firstpage}--\pageref{lastpage}}
\maketitle

\begin{abstract}
When following the growth of structure in the Universe, we propose replacing merger trees with merger graphs, in which \halos\ can both merge and split into separate pieces.  We show that this leads to smoother mass growth and eliminates catastrophic failures in which massive \halos\ have no progenitors or descendants.  For those who prefer to stick with merger trees, we find that trees derived from our merger graphs have similar mass growth properties to previous methods, but again without catastrophic failures.  For future galaxy formation modelling, two different density thresholds can be used to distinguish host \halos\ (extended galactic \halos, groups and clusters) from higher-density \subhalos: sites of galaxy formation.  
\end{abstract}


%%%%%%%%%%%%%%%%% BODY OF PAPER %%%%%%%%%%%%%%%%%%
\section{Introduction}

Structure formation in the Universe is hierarchical in nature, with small \halos\ forming first and merging into ever-larger ones in a process which is usually visualised in the form of a {\bf merger tree}, first popularised by \citet{Lacey93}.

The Sussing Merger Trees project \citep[hereafter SMT13]{Srisawat13} applied 10 different merger tree algorithms on the same dark matter halo catalogue and compared the results.  These all showed a variety of issues characterised by extreme fluctuations in mass along the merger history of a halo, and the occasional abrupt disappearance (i.e.~lack of progenitors) of even quite massive \halos\ looking back in time.  Most of these problems are caused by either: a difficulty in locating the centre about which to grow a halo, leading to occasional ``flip-flops'' in which the halo's centre jumps from one substructure to another; or the disappearance of a satellite subhalo as it passes through another, larger halo, and then its reappearance at a later stage.

A subsequent paper by \citet{Avila14} investigated the extent to which the results of SMT13 were affected by the use of a particular halo finder.  They used a wider range of halo finding techniques and found that, although the detailed statistics of halo growth were altered, the basic problems mentioned above remained.

\begin{figure}
  \centering
  \includegraphics[width=\linewidth]{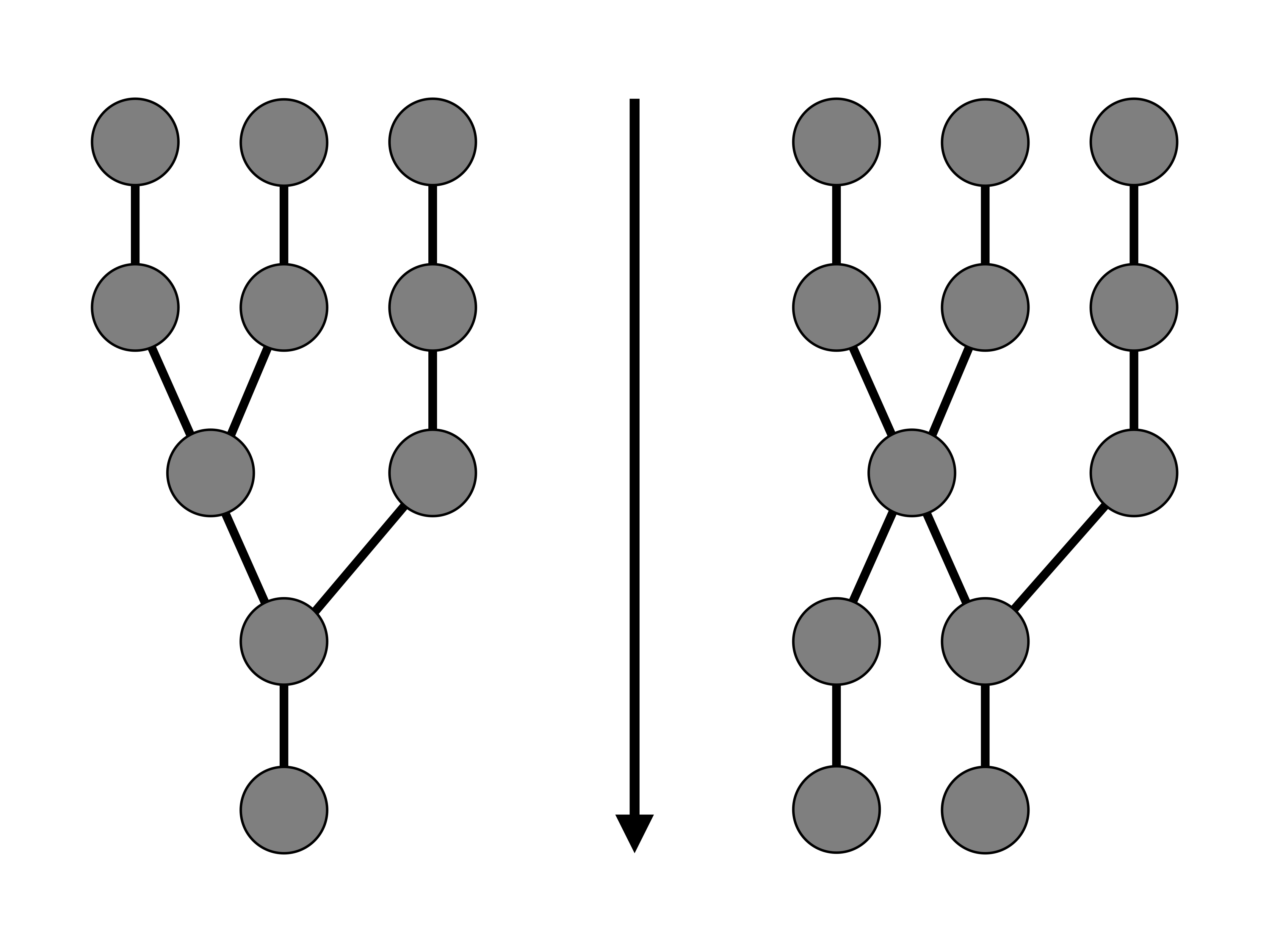}
  \caption{A schematic showing the differences between a merger tree (left) and a merger graph (right). The arrow represents the direction of time with the tree and graph rooted at the present day. The fundamental difference between a merger tree and merger graph is that a merger graph permits \halos\ splitting apart, whereas a tree has a single descendent. This leads to potential extra branches in the graph, branching off towards the present day, as shown here.}
  \label{fig:tree_graph_comp}
\end{figure}

This paper investigates the extent to which these problems can be overcome by the use of {\bf merger graphs} rather than merger trees, in which \halos\ are allowed both to split apart as well as merge together: this fundamental difference between merger trees and the corresponding merger graphs is illustrated in \Fig{tree_graph_comp}.  We will show that they eliminate catastrophic failures (appearance or disappearance of massive \halos) and provide smoother mass growth. Moreover, although that is not our proposed use, we find that such graphs can be split  after construction back into trees, whose properties are more well-behaved than any investigated in SMT13.

One of the main uses of merger trees is to provide a skeleton of the evolution of dark matter \halos, within which to build models of galaxy formation. These models are compromised in existing trees by the flip-flop behaviour mentioned above.  We suggest that the best way to circumvent that is {\bf not} to divide substructures into, as existing methods do, a single {\bf main} halo that is associated with the whole dark matter halo, and other sub-halos which are deemed to be satellites.  Instead we propose that all substructures are treated equally, but with the possibility of deeming one of those substructures to host a {\bf central} galaxy if it is at, or close to, the dynamical centre of the enclosing halo.

We identify \halos\ with a Friends-of-friends (FOF) linking algorithm.  Unlike most existing algorithms, we choose {\bf not} to define a centre about which to grow a spherical overdensity.  That not only introduces the problem of misidentification of main \halos, as mentioned above, but also leads to the possibility of overlapping \halos.  Moreover, we identify substructures using FOF with a smaller linking length, which necessarily leads to the correct nesting of high-density substructures within lower density \halos.  In the context of galaxy formation modelling, the low-density \halos\ would be container \halos\ (clusters, groups, or even extended dark matter \halos\ around individual galaxies) with the higher-density substructures being the localised dark matter \halos\ within which galaxies form.

The paper is constructed as follows.  In the rest of this introduction we explain some of the terminology related to merger graphs, and review previous work.  In \Sec{method}, we describe our \FullName\ (\Name) method for identifying \halos, constructing merger graphs, splitting those graphs back into trees, and nesting of \subhalo{}s.  \Sec{results} first contrasts our merger tree results with those from SMT13, then shows the equivalent plots for merger graphs, before looking at the distribution and dynamics of \subhalo{}s.  Finally, in \Sec{conc}, we present our conclusions and discuss their implications.

\subsection{Terminology}
\label{sec:intro:term}

Where possible we stick to the terminology outlined in SMT13 and expanded upon in \citet[hereafter TOT15]{TOT15}.  We provide a brief glossary of some key terms here.  Firstly,
\begin{shortitem}
\item A {\bf halo} is a dark-matter condensation as returned by a halo-finder.
\end{shortitem}
There are a variety of different halo finding algorithms in use: we describe the ones of relevance to this paper in \Sec{intro:prev} below.
\begin{shortitem}
\item A temporal {\bf merger graph} is a set of ordered halo pairs, ($H_A,\,H_B$), where $H_A$ is older than $H_B$, that represents the growth of structure over cosmic time.
\item Recursively, $H_A$ itself and progenitors of $H_A$ are {\bf progenitors} of $H_B$.  Where it is necessary to distinguish $H_A$ from earlier progenitors, we will use the term {\bf direct progenitor}.
\item Recursively, $H_B$ itself and descendants of $H_B$ are {\bf descendants} of $H_A$.  Where it is necessary to distinguish $H_B$ from later descendants, we will use the term {\bf direct descendant}.
\item Optionally, one of the direct progenitors may be labelled the {\bf main progenitor} -- in this paper we will take that to be the most massive.
\item The longest continuous sequence of main progenitors extending back in time from a given halo is known as the {\bf main branch}.
\item A halo that has no descendants is known as an {\bf end halo}.
\item A temporal {\bf merger tree} is a temporal merger graph in which there is precisely one direct descendant for every halo, except for a single end halo.
\end{shortitem}
The word temporal has been inserted into the definitions above because it is also possible to nest structure at any given epoch into a hierarchy of \halos\ and \subhalos\ of different overdensity.  In this paper we use only two density levels and so we use this restricted terminology:
\begin{shortitem}
\item The spatial nesting of \halos\ is described by a set of ordered halo pairs, $(H_A,H_B)$, where $H_A$ is nested within $H_B$.
\item $H_A$ is a {\bf subhalo} of $H_B$.
\item $H_B$ is the {\bf host halo} of $H_A$.
\item The subhalo closest to the centre of the host halo we will call the {\bf central subhalo} -- note that this differs from TOT15 who use the term main halo.
\end{shortitem}
In addition to the above we use a couple of terms to describe intermediate steps in the graph construction:
\begin{shortitem}
\item A {\bf candidate halo} has been selected for further investigation but is not yet confirmed as a true halo.
  \item When producing trees from our graphs, for comparison with previous work, we often have to divide \halos\ into two or more pieces: we call each of these a {\bf split-halo}.
\end{shortitem}

\subsection{Previous work}
\label{sec:intro:prev}

The first stage in developing robust and complete merger graphs or trees is to identify a halo catalogue.  A comparison of existing techniques was described in the ``\Halos\ Gone MAD'' project \citep{Knebe11}.  Those are primarily based on the FOF algorithm, either in real space or phase space, and/or a spherical overdensity (SO) step \citep{Press74} to grow \halos\ around density peaks. We choose the former method so as to ensure non-overlapping \halos.

As mentioned above, SMT13 and \citet{Avila14} are two studies that look in detail at different merger tree algorithms and compare their properties. To the best of our knowledge there have been no previous published papers on merger graphs of structure formation, except the above-mentioned TOT15 arXiv article that proposes a terminology for merger graphs as part of a longer article on merger tree data format.  We note, however, that \citet{Han2012} produce a subhalo level merger tree and as such implicitly produce a graph at host level when tracking fly-by and ejected \subhalos.

It is not our intention to provide a detailed description of either halo finding or tree-making algorithms in this paper as that is done in great detail in the above-mentioned papers. However, we do make comparison in this paper to much of that previous work, particularly SMT13 whose data we had access to, so we present a brief description of that here.

\subsubsection{The simulation}
\label{sec:intro:prev:sim}

In this work we utilise the simulation produced during SMT13 to directly compare to their results. This cosmological dark matter only simulation was performed using the \textsc{Gadget-3} \textit{N}-body code \citep{Springel2005} with initial conditions drawn from the WMAP-7 cosmology \citep{Komatsu2011}. It has $270^{3}$ particles with a dark matter particle mass resolution of $9.31\times 10^8 h^{-1}\mathrm{M}_{\odot}$. The simulation was performed in a ($62.5 h^{-1}$\,Mpc)$^3$ box, with 62 snapshots covering a redshift range from $z=50$ to $z=0$. We had full access to the output of this simulation in addition to the data produced by the participating merger tree construction algorithms.

\subsubsection{Halo finders}
\label{sec:intro:prev:halofinders}

The Amiga Halo Finder, AHF \citep{Gill:2004aa,Knollmann:2009aa} employs recursively adaptive grids to locate local overdensities in the density field. The identified density peaks are then treated as centres of prospective \halos. After identifying \halos, unbinding is done based on the SO model and halo properties are then calculated based on the particles asserted to be gravitationally bound to their respective density peak. This is the halo finder employed by SMT13 and thus the halo finder we compare directly to.

ROCKSTAR \citep{Behroozi:2013ae} is a halo finder which builds a hierarchy of particles from FOF groups in phase space by progressively and adaptively reducing the linking length. Unbinding is then performed using the full particle potentials and subsequently halo centres are defined by averaging particle positions in a small region close to the phase-space density peak. Although merger tree data was not available for direct comparison, this halo finder is the closest in essence to the approach presented in this paper.

SUBFIND \citep{Springel:2001aa} identifies disjoint, gravitationally bound, locally, overdense regions within an input set of particles, traditionally provided by a FOF group finder. The algorithm then searches for saddle points in the iso-density contours within the global field of the halo to identify substructures. Then the self-boundness condition is asserted so that the particles with positive total energy are iteratively dismissed until only bound particles remain. 

\subsubsection{Merger tree construction algorithms}
\label{sec:intro:prev:mergertreealgs}

Aside from JMERGE (developed as part of SMT13), which estimated the trajectory of objects, all participants in SMT13: MERGERTREE (part of AHF), CONSISTENT TREES \citep{Behroozi:2013ad}, D-TREES \citep{Jiang2014},  HBT \citep{Han2012}, LHALOTREE \citep{Springel:2005ab}, SUBLINK \citep{Rodriguez-Gomez:2015aa}, TREEMAKER \citep{Tweed:2009aa}, VELOCIRAPTOR \citep{Elahi:2019aa,Elahi:2019ab} and YSAMTM \citep{Jung:2014aa}; used particle IDs to link objects between snapshots. To make merger trees, all algorithms choose at most one object as the descendant of an older object, to do this they use a variety of the available halo information. Some only count the fraction of shared particle IDs while others might use extra information such as position, velocity, binding energy, or any other physical properties of the particles or \halos. A summary of these approaches is presented in detail in SMT13. 

Throughout the merger tree analysis presented here we make comparisons to all the available SMT13 algorithms; however, data for comparison to HBT was unavailable for this paper, as such this algorithm is omitted here but presented in full in SMT13: we note that the omission of this algorithm does not effect the result and conclusions of this paper. For clarity, in subsequent analysis we focus on comparison to D-Trees due to it's main branch length performance, being the next best to MEGA. For that reason we present greater detail on D-Trees.

The D-Trees algorithm is designed to work with halo catalogues extracted by the SUBFIND halo finder, like AHF this can lead to missing progenitors and descendants. To combat this D-Trees allows for descendants to be identified multiple snapshots later, with a free parameter $N_{\mathrm{step}}$ defining how many snapshots are used for the descendent search; for SMT13, $N_{\mathrm{step}}=5$ was used. 

These descendants are identified by finding a most bound "core" of particles from the particles in all progenitors in snapshot A. A number of linking particles is then defined from the group of progenitors, a descendant is then any halo in snapshot B (where snapshot B is within snapshots A$+1$ to A$+N_{\mathrm{step}}$) which shares at least 1 of these linking particles. The descendent at snapshot B is then the candidate descendent with the maximum fraction of linking particles. This can yield up to $N_{\mathrm{step}}$ descendants. To identify the descendent used in the merger tree, the main progenitor of each is found: if this main progenitor is the group at snapshot A for more than one of the descendants, then the earliest descendent is taken; if it is the main progenitor for only one descendent, then this descendent is taken; or if no such descendent exists then the earliest descendent is taken regardless. 

We note that some tree construction algorithms were designed to work with bespoke halo finders and, as such, the SMT13 comparison may not show them in their best light.  This was investigated in \citet{Avila14}.  Unfortunately, we do not have access to data from that paper and so cannot add those results to our plots, for which the reader will have to refer back to the original paper.  We note, however, that the improvement is modest compared to the results of our own \Name\ algorithm that we describe below.

\section{Method}
\label{sec:method}

In \Sec{method:halos}, we describe our \FullName\ (\Name) method for identifying \halos: because of our desire to have distinct, non-overlapping \halos, that will be based solely on the Friends of Friends \citep[FOF, ][]{Davies85} algorithm together with a binding energy check.  We then, in \Sec{method:graphs}, describe the manner of linking those \halos\ into merger graphs.  To facilitate comparison with previous work, we describe in \Sec{method:trees} a method of splitting those graphs up into trees.  Finally, in \Sec{method:nesting}, we introduce the concept of spatial nesting of \halos\ of differing overdensity (i.e. FOF linking length).

Throughout this paper we will use the same input simulations as in SMT13 and contrast our results to those from the various algorithms used in that paper.

\subsection{Identification of \halos}
\label{sec:method:halos}

We will first describe the identification of \halos\ in configuration space and then extend that to phase space, which we will show is better able to separate 'fly-by' \halos\ that have temporarily merged and eliminate spurious \halos.
A flow-chart summarising the halo finding algorithm is shown in \Fig{haloflow} in \App{flows}.

\subsubsection{Spatially-defined (configuration space) \halos}
\label{sec:method:halos:real}

We generate candidate \halos\ by linking together all particles with separations less than
\begin{equation}
\ell_{s} = \alpha_{s} \bar{d},
\label{eq:spatial_ll}
\end{equation}
where $\bar{d}$ is the mean inter-particle separation and $\alpha_{s}=0.2$; this corresponds roughly to an overdensity with respect to the mean of 300-550 \citep{More11}. Candidate \halos\ are required to contain a particle number $N_p\geq10$.  This minimum particle number is rather small and means that many of the smallest candidate \halos\ will be spurious: however we want to maximise the chances that a larger halo, once it has formed, will survive until the present day (albeit that it may merge with a more massive halo).

For a candidate halo to make it into the final halo catalogue, we require one of the following to be true, whose motivation is
explained below: {\bf either}
\begin{shortitem}
\item The number of particles, $\Np \geq 20$ {\bf and}
\item The total energy of the halo, $E\leq0$;
\end{shortitem}
    {\bf or}
\begin{shortitem}
\item The halo has 10 or more particles in common with a halo in the previous snapshot.
\end{shortitem}
This ensures that each halo is at least 20 particles in size when it first appears, although it may fluctuate below that in its subsequent evolution.

The energy of a halo, $E$, is defined as
\begin{equation}
E=\mathrm{KE}+\mathrm{GE},
\end{equation}
where
\begin{equation}
\mathrm{KE}={1\over2}M\sigma_v^2\equiv\sum_{i=1}^{\Np}\frac{1}{2}m_i|\mathbf{v}_i-\langle\mathbf{v}\rangle|^2
\end{equation}
and
\begin{equation}
\mathrm{GE} = -\sum_{i=1}^{\Np}\sum_{j=1}^{i-1}\frac{G m_i m_j}{({r_{ij}^2+s^2)^{1/2}}}.
\end{equation}
Here the sums run over the particles in a halo, $i=1\ldots\Np$; $m_i$ is the mass of particle $i$ and $M$ the total mass of the halo; $\mathbf{v}_i$ is the particle velocity, $\langle\mathbf{v}\rangle$ the mean velocity of the halo, and $\sigma_v$ its (3-D) velocity dispersion; $G$ is the gravitational constant; $r_{ij}$ is the separation of particles $i$ and $j$; and $s$ is the softening of the simulation.

\begin{figure}
  \centering
  \includegraphics[width=\linewidth]{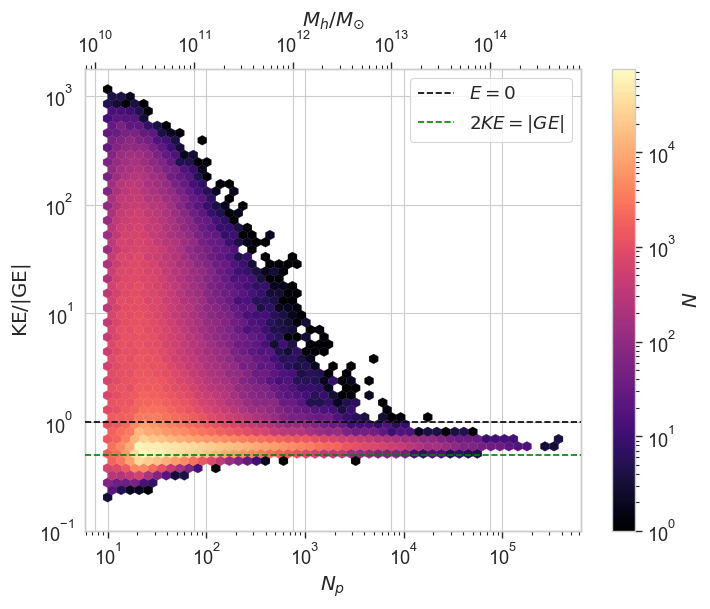}
  \caption{The energy distribution of spatially-defined \halos\ with the 'realness' criterion applied as a function of their mass.  The colour of the markers signifies the density of points in that region of the distribution. Where $N_{p}<20$, \halos\ are included only if they track the evolution of a halo which at a previous time had $N_{p}\geq 20$ (i.e. was deemed real).}
  \label{fig:spatialpersistcutenergy}
\end{figure}

\Fig{spatialpersistcutenergy} shows the value of $E$ for each of our spatially-defined \halos.  The lower dashed line shows the relation expected for an isolated, virialised halo (whose size is much larger than the softening).  The modal relation for our \halos\ lies slightly above this line principally because they are not isolated from their surroundings but are exchanging particles across the nominal outer density contour of the halo.

It is clear that a substantial fraction of \halos\ have positive energy.  At low particle number, \Np, most of these are spurious associations of particles that have broken away from a larger halo and which are either subsequently reincorporated or rapidly disperse.  It would be possible to weed these out based upon their future behaviour, and indeed we have a method to do this that we included in an earlier draft of this paper; however, we discovered that the phase-space linking described below does a much better job and so we decided to omit that more complicated spatial algorithm.

There are also \halos\ with very high particle number, $\Np > 1000$, even one with $\Np > 10\,000$, that have positive energy: these are merging systems whose orbital energy dominates over the binding energy of the merging systems.

Note that we do not consider whether individual particles are bound to the halo, which is at the heart of many other methods.  That means that the outer parts of our FOF \halos\ may well contain particles that are not bound (and, conversely, there may be particles that lie outside the FOF halo that are gravitationally bound to it). There are two reasons for this: firstly, unbinding is an ill-defined and costly procedure; and secondly, it can lead to disruption. and hence disappearance from the graphs, of merging \halos.  We find that the main positive effects of particle unbinding, namely identifying spurious \halos, or removing 'fluff' from the outskirts of \halos, can be accomplished by moving to phase-space linking, as described in the next subsection.

\subsubsection{Phase Space}
\label{sec:method:halos:phase}

To reduce the number of massive merging \halos\ with $E>0$ which are in fact genuine interacting systems, while also reducing the number of temporary particle associations, we move to a phase space halo definition. We do this by initially applying the spatial metric
\begin{equation}
\frac{r_{ij}^2}{\ell_{s}^2} \leq 1,
\end{equation}
to produce candidate \halos, which are then individually passed through the phase space metric
\begin{equation}
\frac{r_{ij}^2}{\ell_{s}^2} + \frac{v_{ij}^2}{\ell_{v}^2} \leq 2,
\end{equation}
where $r_{ij}$ is the particle separation in real space, $v_{ij}$ is the particle separation in velocity space, $\ell_{s}$ is the previous spatial linking length (\Eq{spatial_ll}) and $\ell_{v}$ is the velocity space linking length.

We define an adaptive velocity space linking length derived from the virial theorem for an isolated halo of mass $M$, with particle mass $m_p$, number of particles $\Np$, and overdensity $\Delta$ relative to the mean density $\bar{\rho}$. We start with the virial theorem for an isolated isothermal sphere\footnote{In reality there is no such thing, but we only need an approximate value for the default linking length.  For the same reason, we ignore the distinction between $\Delta$ and $\Delta+1$.}
\begin{equation}
2\times\frac{1}{2}M\sigma_v^2+\frac{GM^2}{2r}=0,
\end{equation}
where $r$ is the radius of a sphere of density $\Delta\bar{\rho}$ and mass $M$,
\begin{equation}
  r = \left(\frac{3M}{4\pi\Delta\bar{\rho}}\right)^{1/3}.
\end{equation}
Hence
\begin{equation}
  \sigma_v^2=\frac{G}{2}\left(\frac{4\pi\Delta\bar{\rho} M^2}{3}\right)^{1/3}
  =\left(\frac{\Delta\bar{\rho}\pi G^3(m_{p}\Np)^2}{6}\right)^{1/3}.
\end{equation}
From this characteristic velocity we then define the velocity space linking length,
\begin{equation}
\ell_{v}=\alpha_{v}\left(\frac{\Delta\bar{\rho}\pi G^3(m_{p}\Np)^2}{6}\right)^{1/6},
\label{eq:vellinkl}
\end{equation}
where we have introduced the free parameter $\alpha_v$ whose value we vary as described below.

\begin{figure}
  \centering
  \includegraphics[width=\linewidth]{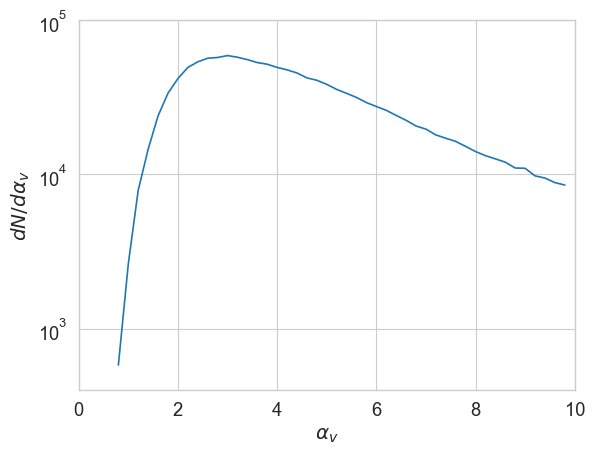}
  \caption{The distribution function of velocity linking length coefficients, $\alpha_v$, that \halos\ exit the phase space iteration with after achieving $E\leq0$. Most \halos\ are bound with $\alpha_v=10$ and so are omitted from the plot, as are those few that remain unbound for all $\alpha_v\geq0.8$.}
  \label{fig:alphavhist}
\end{figure}

Using too small a value of $\alpha_v$ can lead to some of the smaller genuine \halos\ disappearing from the catalogue, while using too large a value fails to separate some obvious mergers.  Additionally, as mentioned above, using a single linking length leads to a range of recovered overdensities for FOF \halos. For these reasons, we nominally set $\Delta=200$ and loop over successively smaller values of $\alpha_v$ from 10 down to 0.8, exiting the loop and saving the properties of the halo(s) if at any stage $E<0$.  Should any \halos\ break into 2 or more pieces then each is treated in the same manner.  Should a halo reach $\alpha_v=0.8$ with $E>0$ it is only included in the final catalogue if it is required for another halo's persistence as described in \Sec{method:halos:real}.

\begin{figure}
  \centering
  \includegraphics[width=\linewidth]{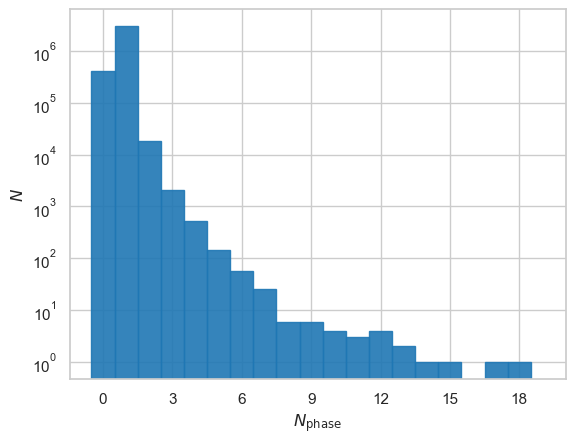}
  \caption{The number distribution of phase space counterpart \halos\ for each spatially-defined candidate halo. A phase space number of 0 corresponds to a spurious spatial halo which has no phase space counterpart.}
  \label{fig:numphasehalos}
\end{figure}

\begin{figure}
  \centering
  \includegraphics[width=\linewidth]{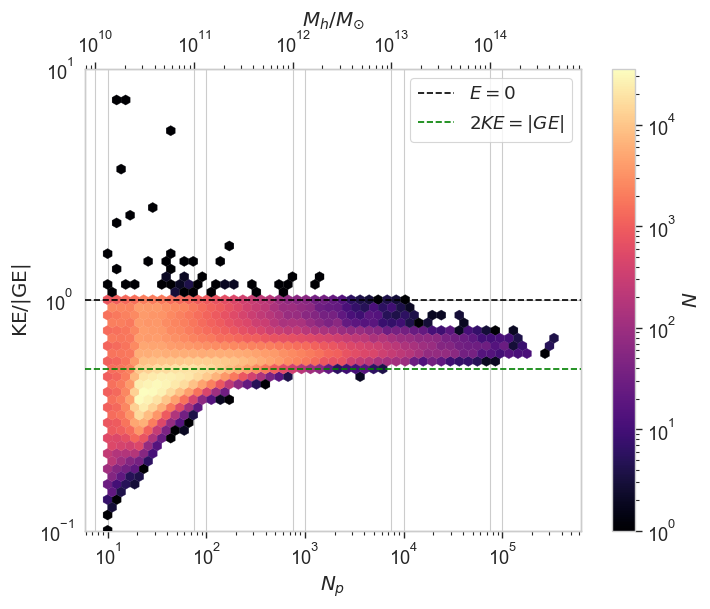}
  \caption{The energy distribution of all host \halos\ at all time steps after applying phase-space splitting. As with \Fig{spatialpersistcutenergy} the colour of the markers signifies the density of points in that region of the distribution and $N_{p}<20$ \halos\ are included only if they track the evolution of a halo which at a previous time had $N_{p}\geq 20$.}
  \label{fig:phaseenergy}
\end{figure}

The final $\alpha_{v}$ distribution is shown in \Fig{alphavhist} for those \halos\ which exit the iteration having reached $E\leq0$, the number of \halos\ that result from each spatial counterpart is shown in \Fig{numphasehalos}, and the energy of the surviving \halos\ is shown in \Fig{phaseenergy}.  As expected from \Fig{spatialpersistcutenergy}, the vast majority (about 80\,\%) of \halos\ have negative energy even when all particles are included and are thus unaffected by this move to phase space.  However, 8\,\% only attain negative energy with a reduced velocity linking length, while 12\,\% of all candidate \halos\ are excluded because their particle count drops below 10 whilst their energy is still positive. Somewhat surprisingly, only 116 out of the original population of almost 1 million candidate \halos\ with $E>0$ survive with 10 or more particles and positive energy at the end of this process -- it is extremely efficient at removing rogue \halos.  Of those 116, 59 are retained because they are the descendant of a real halo from the previous snapshot.

\begin{figure}
  \centering
  \includegraphics[width=\linewidth]{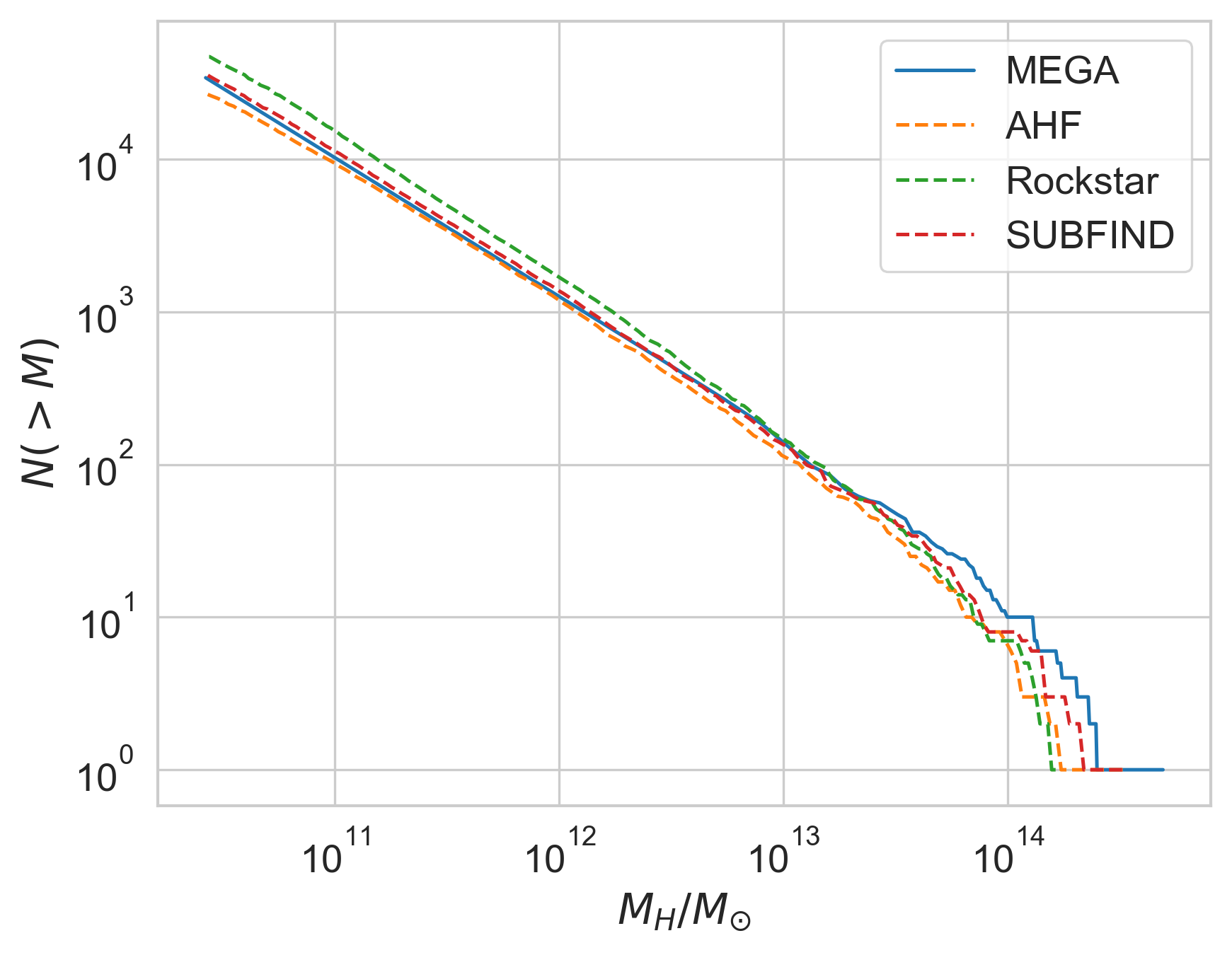}
  \caption{A comparison of the present day ($z=0$) halo mass function for the halo catalogue produced from this simulation by the \Name\ algorithm, Amiga Halo Finder (AHF), Rockstar and Subfind.}
  \label{fig:massfunc}
\end{figure}

\Fig{massfunc} shows the halo mass function at $z=0$ in comparison to several other halo finders.  Our method produces very similar halo numbers to \subfind\ and \AHF\ at low masses, but our FOF method returns higher masses for the largest \halos.  We stress, once again, that out \halos\ are not directly comparable to those in SMT13 because we do not require individual particles to be bound.  The justification for this looser definition comes in the improved behaviours of the resultant graphs and trees described in \Sec{results} below.

\subsection{Construction of merger graphs}
\label{sec:method:graphs}

\begin{figure}
  \centering
  \includegraphics[width=\linewidth]{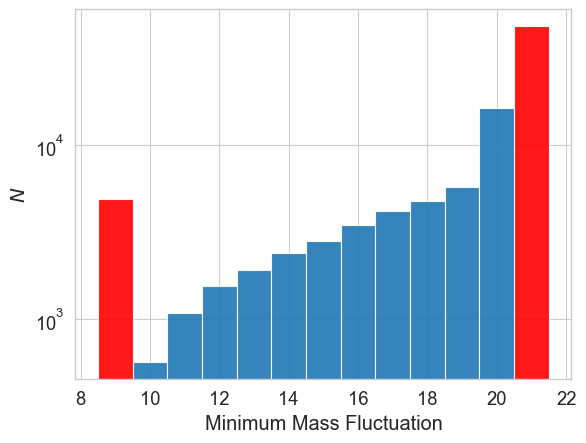}
  \caption{The distribution of minimum mass a halo reaches in its subsequent evolution once it has initially reached a particle mass of $N_p=20$ or greater. \Halos\ included in the bin with mass less than $<10$ are those \halos\ which drop below the 10 particle limit and are no longer tracked. The \halos\ in the $N_p=20$ bin are those that never drop below 20 particles once they reach that mass; those in the $N_p=21$ bin are \halos\ which initially have a mass greater than 20 particles and never have $N_p\leq20$ particles in their subsequent evolution.}
  \label{fig:minmassfluc}
\end{figure}

Given a halo catalogue, the construction of merger graphs in our method is elementary: we simply link together \halos\ in adjacent snapshots (output times) that have 10 or more particles in common.   Recall that we require that all \halos, when they first appear, are at least 20 particles in size, but may fluctuate down to 10 particles in their subsequent evolution.  The distribution of minimum mass for a halo over its lifetime is shown in \Fig{minmassfluc}. About a quarter of \halos\ fluctuate below the 20 particle limit but only a small fraction (about 1 in 20) drop below 10 particles and so give truncated branches that don't reach the present day.

\subsection{Splitting merger graphs into merger trees}
\label{sec:method:trees}

To facilitate comparison with previous work, we introduce a method to split graphs back into merger trees.  A flowchart summarising this procedure is shown in \Fig{splitflow} of \App{flows}.

Starting with the penultimate snapshot and working backwards in time, we consider the descendants of each halo.  If there is more than one descendant then we treat the halo as a candidate for splitting.  The rationale is that the halo is a combination of two or more smaller \halos\ that are dynamically distinct but which have temporarily come within the FOF linking length of one another before subsequently splitting apart into multiple descendent \halos.  We define the potential dynamically distinct split-\halos\ by the particles that they have in common with each descendant and test their reality with the following constraints: that the split-halo is the most massive candidate split-halo, that it has negative energy, or that it occupies a distinct region in phase space relative to the most massive split-halo.  This overlap condition is given by the following expression, in a similar approach to the method implemented by \citet{Han2018} to identify trapped sub\halos,
\begin{equation}
  \label{eq:splithalosep} \frac{|\langle\mathbf{r}\rangle_1-\langle\mathbf{r}\rangle_2|}{\Rhalo_{,1}+\Rhalo_{,2}}+\frac{|\langle\mathbf{v}\rangle_1-\langle\mathbf{v}\rangle_2|}{\sigma_\mathrm{v,1}+\sigma_\mathrm{v,2}} \geq0.85,
\end{equation}
where $\langle\mathbf{r}\rangle$ and $\langle\mathbf{v}\rangle$ are the mean positions and velocities, $\Rhalo$ and $\sigma_v$ are the root-mean-square radii and velocities, and the subscripts 1 and 2 refer to the most massive split-halo and the candidate split-halo.

\begin{figure}
  \centering
  \includegraphics[width=\linewidth]{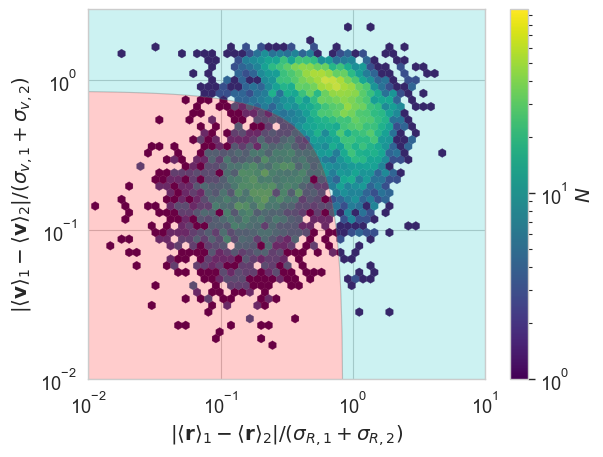}
    \caption{The separation of splitting candidate \halos\ (see \Sec{method:trees}) in real and velocity space where the candidate that is not the most massive split-halo has $E>0$. The boundary between coloured regions represents the condition we use (\Eq{splithalosep}) to distinguish single \halos\ (red) from genuine overlaps (cyan).}
  \label{fig:splithalosep}
\end{figure}
The choice of coefficient in \Eq{splithalosep} is motivated by \Fig{splithalosep} which shows the separations of split-halo pairs in position and velocity space.  Those at large separations in this diagram are genuine overlapping \halos\ that have failed to be separated by the phase-space halo identification method: these we accept as genuine distinct \halos.  The long tail of low-separation candidates, however, are most likely not distinct \halos\ but simply random associations of particles that come together to form the distinct descendant: for these, we omit the candidate split-halo from our catalogue.

Once we have identified all genuine split-\halos\ for any given halo, we allocate each a fraction of the original halo mass in proportion to the number of particles that it contains (for this purpose, the particles from rejected split-halo candidates are associated with the most massive split-halo, with which they overlap).  Thus the mass of the graph halo is equal to the sum of masses of all the tree halos that it is split into.

\begin{figure}
  \centering
  \includegraphics[width=\linewidth]{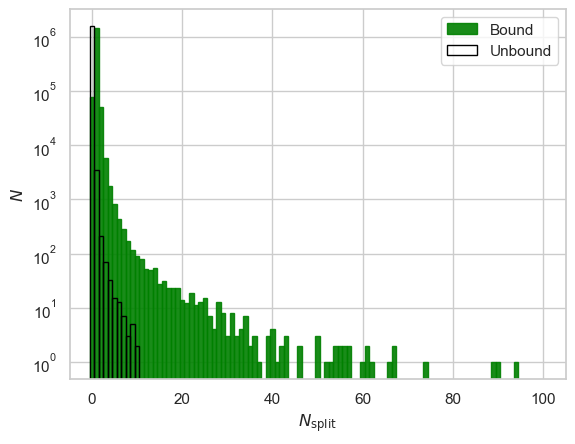}
  \caption{Histograms of the number of split-\halos\, distinguishing between the number of instances of $N_{\mathrm{split}}$ split-\halos\ that have negative and positive energy.}
  \label{fig:number_split_halos_all}
\end{figure}

In this way, after working back in time to the earliest \halos, the graph has become teased apart into separate trees, one for each halo in the graph at the final snapshot.  The resultant number of \halos\ split from all candidate \halos\ is shown in \Fig{number_split_halos_all}.  There are two histograms shown here: one for the number of bound split-\halos\ that each halo is divided into, and another for the number of unbound split-\halos, the larger numbers of unbound \halos\ are isolated to split-\halos\ with a large total $N_{\mathrm{split}}$. The main thing to take away is that even having a single unbound split-halo is very rare (about 1 in $10^3$) and having multiple ones even more so.  To create a tree structure, some \halos\ need to be split into as many as 100 pieces, but these are almost all bound objects.

\begin{figure}
  \centering
  \includegraphics[width=\linewidth]{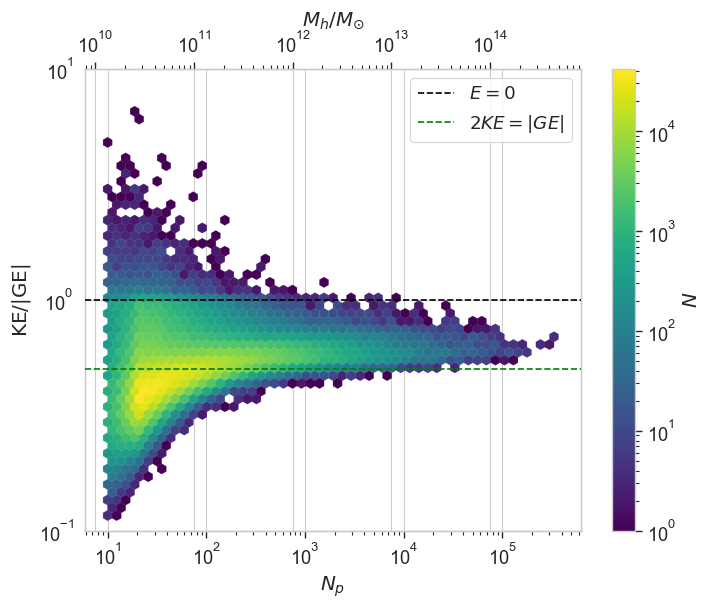}
  \caption{The energy distribution of all \halos\ at all time steps after splitting graphs into their constituent trees, {\it using only those particles in common between split-halos and their descendants}.}
  \label{fig:splitreal}
\end{figure}

To further check that our algorithm does not simply tear apart real objects, we show the energy distribution of these split-\halos\ in \Fig{splitreal}.  Note that some \halos\ with $N_{p}>100$ that were previously bound (and even one with $\Np>1000$) have now show up as having $E>0$ -- that is because we are using in the energy calculation only those particles in common with the descendant.  Of these split-\halos\ with $E>0$, the vast majority have a real ($E<0$) progenitor.  The remaining 18 exist for precisely 2 snapshots with $E>0$ before linking to a real progenitor.

Overall, the splitting procedure seems to have done a reasonable job (and we will see in \Sec{results:trees} below that the resultant tree has properties that match, or outperform, other methods).

\subsection{\Subhalos: spatial nesting}
\label{sec:method:nesting}

The standard linking length of 0.2 times the mean interparticle separation gives \halos\ that roughly correspond to the
virialised region.  The edges of such a region tend to be somewhat irregular and a certain amount of exchange of
material with the surroundings is expected -- hence our use of merger graphs rather than merger trees.  In the
current-day Universe, at least, these tend to correspond to groups and clusters of galaxies: for that reason, we regard
them as container vessels within which higher-density galactic \halos\ will reside.

The overdensity of a halo (or that region of a halo) within which galaxies form is uncertain and probably
redshift-dependent, but will nevertheless be higher than the virial overdensity.  Within this paper, we simply choose a
linking length of 0.1 times the mean inter-particle separation to define such high-density \subhalos.  Exactly as for the host \halos, we select candidate \subhalos\ in configuration space, then try successively smaller velocity linking lengths in phase-space until the energy becomes negative, $E<0$.

With this definition, all \subhalos\ are distinct from one another (no particles in common) and are spatially nested within the corresponding host halo.

\section{Results}
\label{sec:results}

We begin our results in \Sec{results:trees} by looking at the properties of merger trees.  That is not because we see merger trees as the way forwards, but because we want to compare to previous work.  We will show that our trees show similar mass fluctuations to other FOF-based methods in the SMT13 tests, but show fewer catastrophic failures and hence longer main branch lengths.

We then look in \Sec{results:graphs} at the properties of the graph as a whole.  These show, as expected, smoother mass growth than for the trees.

Finally, in \Sec{results:subs} we consider the properties of subhalos, defined as higher density substructures within \halos.

\subsection{Merger trees}
\label{sec:results:trees}

Although the main purpose of this paper is to investigate the properties of merger graphs, we explained in \Sec{method:trees} how to split the graphs back up into trees.  This allows for comparison with previous work and in particular the study of SMT13: we reproduce the plots from that paper below.

We note two differences between our work and most of the SMT13 trees that should be borne in mind when making this comparison.  Firstly, we are defining \halos\ on the basis of FOF groups, rather than spherical overdensity -- our \halos\ are therefore irregular in shape and their mean density is rather poorly defined; to compare the results presented here to FOF defined merger trees see the results of \citet{Avila14}, particularly figures 3, 5, 7 and 8.  Secondly, we require \halos\ to have a minimum of 20 particles when they are first created, but we allow them to fluctuate in mass down to 10 particles in their subsequent evolution -- this has the effect of reducing the number of low mass \halos\ that are brought into existence but then disappear with no descendants.

\subsubsection{Main branch length}
\label{sec:results:trees:mainbranchlength}

\begin{figure}
  \centering
  \includegraphics[width=\linewidth]{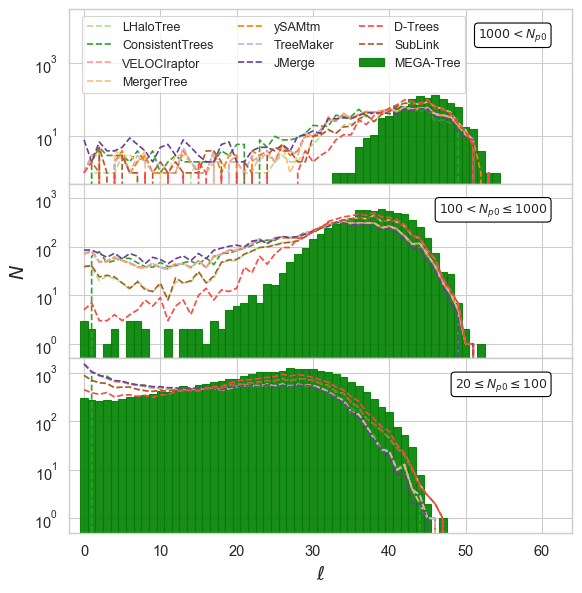}
  \caption{The main branch length of all trees rooted at the present day ($z=0$).  Here $\ell$ is the number of snapshots that a tree's main branch can be traced until it disappears. The upper, middle and lower panels represent mass bins of $\Np > 1000$, $1000\geq\Np>100$ and $100\geq\Np\geq20$ respectively. The \Name\ main branches are defined such that they are rooted at and end on \halos\ with $\Np\geq 20$; in between they are allowed to fall below the 20 particle threshold but no lower than 10 particles.}
  \label{fig:treemainbranch}
\end{figure}

The first test of the quality of the trees that we consider is the main branch length: i.e.~the number of simulation timesteps (snapshots) that a halo can be traced back into the past until it disappears.  \Fig{treemainbranch}\ shows that, grouped by halo mass (particle number) at the current day, and contrasted with the results from SMT13.

The most important thing to note is the absence of a tail extending to short main branch lengths in the upper panel and the reduced tail in the middle panel.  That is to say, all massive \halos\ have existed for an extended period of time, as one might expect.  That is not true of the trees produced by the SMT13 study.  In the middle histogram, for example, those show multiple \halos\ with more than 100 particles that come into existence only on the final snapshot.  More worryingly, there are \halos\ with more than 1000 particles at the present day that exist for only one or two snapshots.  Most likely these are pathological cases whereby a large halo has split into two pieces, only one of which can be linked into the tree.

Even if we exclude these outliers, the median branch length is higher in our trees derived from the merger graphs: the maximum branch length is similar to that in SMT13 but the distribution is narrower.  For the lower panel only, part of the improvement is due to our allowing \halos\ to fluctuate below 20 particles; we show the equivalent plots with a hard 20 particle lower limit in \App{10part}.

\subsubsection{Number of progenitors}

\begin{figure}
  \centering
  \includegraphics[width=\linewidth]{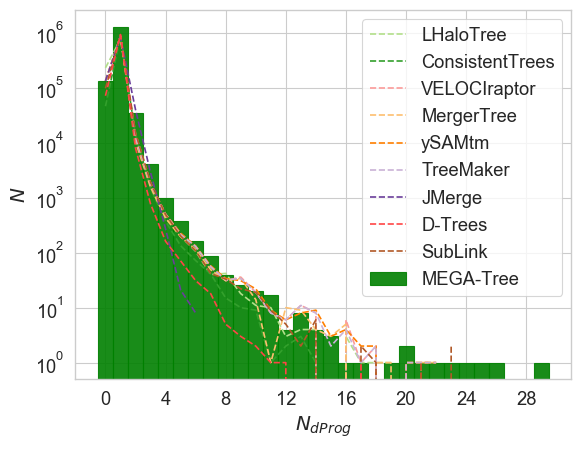}
  \caption{Histograms showing the number of direct progenitors of each halo at all timesteps. For \Tree\ we include only \halos\ with $\Np\geq 20$, for comparison with previous work.}
  \label{fig:treeproghist}
\end{figure}

\Fig{treeproghist} shows the number of direct progenitors for each halo in the trees.  For \Tree\ we restrict our analysis to \halos\ with $\Np\geq20$ to facilitate comparison with the SMT13 results.  It can be seen that \Tree\ is roughly consistent with those results but at the upper end of the progenitor number.  The reason for this is that our looser definition of \halos, based on FOF rather than spherical overdensity, leads to slightly more low mass \halos\ (as can be seen in the lower panel of \Fig{treemainbranch}).

\subsubsection{Mass growth}

\begin{figure}
  \centering
  \includegraphics[width=\linewidth]{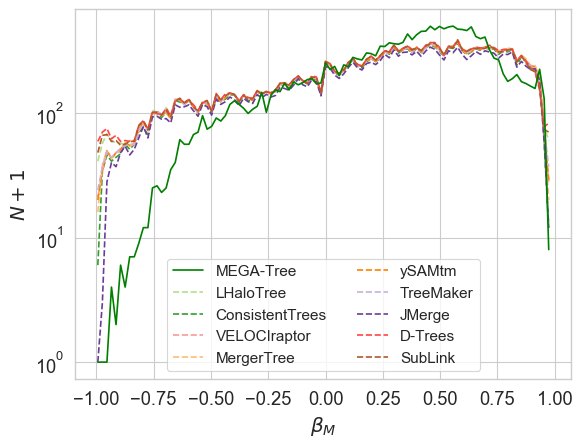}
  \caption{The distribution function of logarithmic mass growth along the main branches of all trees. This represents the slope of the mass evolution between all pairs of halo and descendant in a main branch, each with $\Np\geq 1000$.}
  \label{fig:treelogm}
\end{figure}

We next examine the mass growth along the main branch of the merger trees. \Fig{treelogm}\ shows the slope of the logarithmic mass growth between any two main branch \halos\ with $N_{p}>1000$ in adjacent snapshots. This statistic is defined to lie in the range $-1\leq \beta_m \leq 1$ with
\begin{equation}
    \beta_m(k-1, k) = \frac{\arctan\left(\alpha_{m}(k-1, k)\right)}{\pi/2},
\end{equation}
where the logarithmic mass growth is defined as
\begin{equation}
\frac{d \log(M)}{d \log(t)} \approx \alpha_m(k-1, k) = \frac{(t_{k}+t_{k-1})(M_{k}-M_{k-1})}{(t_{k}-t_{k-1})(M_{k}+M_{k-1})}.
\label{eq:logmassgrowth}
\end{equation}
Thus $\beta_m\sim-1$ represents a catastrophic decrease in mass, and $\beta_m\sim+1$ an abrupt increase in mass.

This statistic shows that is much more common for \halos\ in \Tree\ to experience modest mass growth between snapshots than those from SMT13 trees, and that the fraction of \halos\ that decrease in mass is lower. These differences between SMT13 algorithms and \Tree\ match those found \citet{Avila14} who showed that they can almost entirely be attributed to the difference in halo definition: FOF here compared to spherical overdensity in SMT13. Another difference, which is not simply due to the use of FOF, is that the number of catastrophic changes in mass $|\beta_m|\geq0.75$, is reduced relative to SMT13 algorithms.

\begin{figure}
  \centering
  \includegraphics[width=\linewidth]{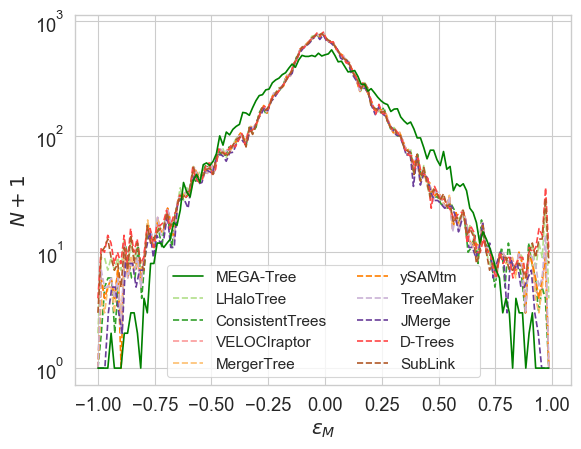}
  \caption{The mass fluctuation of all sets of 3 consecutive temporally linked \halos\ in the main branch of a tree, each with $\Np\geq 1000$.}
  \label{fig:treemassfluc}
\end{figure}

Another method to categorise the mass growth is the fluctuation in the rate of mass growth: this is presented in \Fig{treemassfluc} for all successive main branch \halos\ with $N_{p}>1000$. The mass fluctuation is given by
\begin{equation}
\epsilon_{m} = \frac{\beta_m(k,k+1) - \beta_m(k-1,k)}{2}.
\label{eq:massfluc}
\end{equation}
At first sight the difference between Figures~\ref{fig:treelogm} and \ref{fig:treemassfluc} seem odd: the former shows a greater tendency for positive mass growth in \Tree\ than in SMT13, but a broader spread of fluctuations in the rate of growth.  However, we have checked that this behaviour is real, and a similar effect was again observed in \cite{Avila14} for FOF-defined \halos\ as compared to the spherical overdensity defined \halos\ in SMT13.

So in summary, our \Tree\ trees show similar fluctuations in mass to previous algorithms, but far fewer catastrophic failures, i.e.~sudden appearance or disappearance of massive \halos.

\subsection{Merger graphs}
\label{sec:results:graphs}

We now investigate the main focus of this work, merger graphs. Where possible we compare statistics with those produced by trees, both in this work and in SMT13.

\subsubsection{Main branch length}

\begin{figure}
  \centering
  \includegraphics[width=\linewidth]{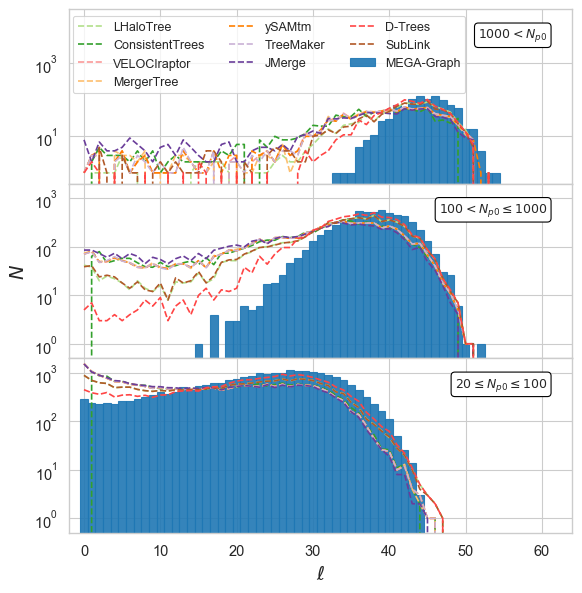}
  \caption{The main branch length of all graphs rooted at the present day ($z=0$). As in Fig.\ref{fig:treemainbranch}, $\ell$ is the number of steps or generations possible 'walking' down the tree, with the lower, middle and upper panels representing mass bins of $20\leq \Np \leq 100$, $100< \Np \leq 1000$ and $1000\leq \Np$ respectively. Here, the \Name\ main branches are defined such that they are rooted at generations of the graph with at least one halo with $M\geq 20$ and end on \halos\ with $M\geq 20$; in between \halos\ are allowed to fall below the 20 particle threshold but no lower than 10 particles.}
  \label{fig:graphmainbranch}
\end{figure}

Once again the first quality of graphs we measure is the main branch length, grouped by halo mass (particle number) at the current day, and contrasted with the results from SMT13 in \Fig{graphmainbranch}.  

As with the trees, the median branch length is higher than that in SMT13 with a narrower distribution and a similar maximum branch length.  This can be attributed to our merger graphs being better behaved at low particle numbers which is again evident in the lower panel of \Fig{graphmainbranch} with no sharp upturn at short branch lengths relative to the upturn on SMT13 trees at the same point.

More importantly, catastrophic failures in the evolution of \halos\ with $\Np>100$ are completely eliminated.  Unlike for trees, the graph does not require decisions to be made about which descendants to keep and so there is no possibility that halo will become detached from the graph.

\subsubsection{Number of progenitors and descendants}

\begin{figure}
  \centering
  \includegraphics[width=\linewidth]{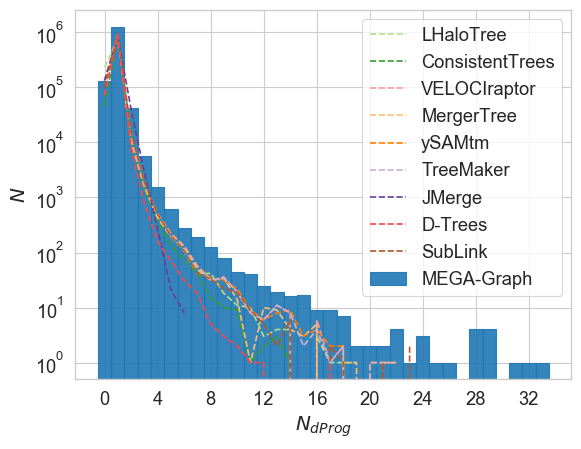}
  \caption{Histograms showing the number of direct progenitors of each halo in all timesteps. Once again, for the SMT13 algorithms this includes all \halos\ within their trees; for \Graph\ this includes all \halos\ with $\Np\geq 20$.}
  \label{fig:graphproghist}
\end{figure}

The number of direct progenitors of \Graph\ \halos\ is shown in \Fig{graphproghist}.  Unsurprisingly, this is similar to the result for \halos\ in the \Tree, shown in \Fig{treeproghist}.

\begin{figure}
  \centering
  \includegraphics[width=\linewidth]{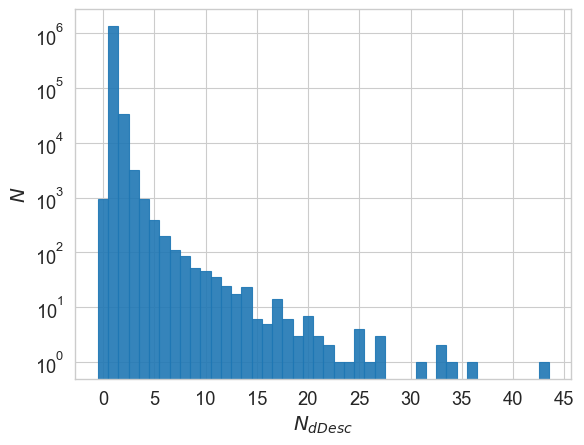}
  \caption{Histogram showing the number of direct descendants with $\Np\geq 10$ for all \Graph\ \halos\ in all snapshots.}
  \label{fig:graphdeschist}
\end{figure}

More interesting is the number of direct descendants, shown in \Fig{graphdeschist}.  For a tree this is mostly 1, except for the rare occasions when a halo goes missing, in which case it is 0.  Notice that the \Graph\ algorithm has not eliminated 0 descendants (end \halos) entirely: these are those \halos, as discussed in \Sec{method:graphs} and shown in \Fig{minmassfluc}, whose mass eventually dwindles below a particle number of 10.

\subsubsection{Smoothness of mass growth}

\begin{figure}
  \centering
  \includegraphics[width=\linewidth]{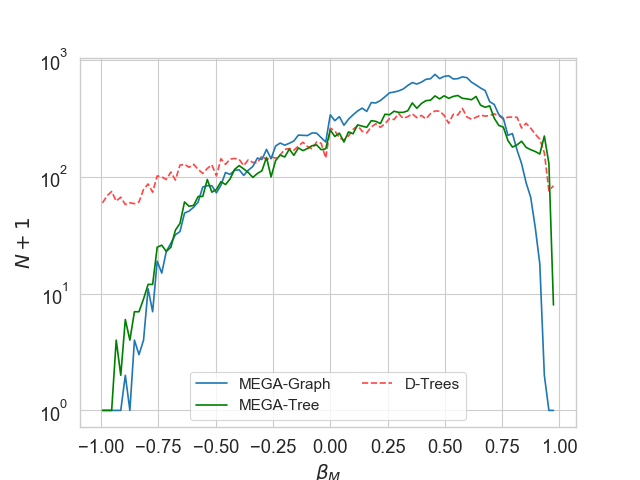}
  \caption{The distribution of logarithmic mass growths between graph generations compared to the main branch of \Tree\ and those of the SMT13 algorithm D-Trees. This plot represents the slope of the mass evolution between all pairs of generations with a total number of particles $\Np\geq 1000$. For the trees this is all pairs of halo and descendent in a main branch both with $\Np\geq 1000$.}
  \label{fig:graphlogm}
\end{figure}

\begin{figure}
  \centering
  \includegraphics[width=\linewidth]{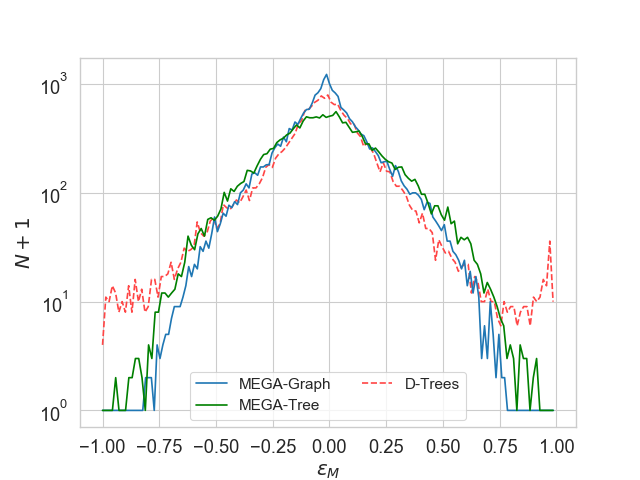}
  \caption{The mass fluctuation of all sets of 3 consecutive graph generations with $\Np\geq 1000$. Again we present \Tree\ and D-Trees for comparison where 3 consecutive main branch \halos\ have $\Np\geq 1000$.}
  \label{fig:graphmassfluc}
\end{figure}

It is when we look at the smoothness of the mass growth that the value of graphs over trees becomes clear.  \Fig{graphlogm} shows that the mass growth is much more strongly biased towards positive values even than the \Tree{}s and that the abrupt increases and decreases of mass have been entirely eliminated.  Similarly, in \Fig{graphmassfluc}, we see that mass fluctuations are smoother in the \Graph{}s than in any tree.  While not entirely surprising, this does reinforce the idea that graphs provide a more complete framework for following the growth of structure than do trees.

\subsection{Substructure}
\label{sec:results:subs}

In this section we look at the properties of \subhalos\ within our host \halos.  As a reminder, for the purposes of this paper, \subhalos\ are simply defined to be \halos\ defined in exactly the same way as for the host \halos, but with linking lengths in both position and velocity corresponding to 8 times the overdensity.

We define the central \subhalo\ to be the the one whose centre of mass is closest to the centre of mass of the enclosing host halo -- in 9 out of 10 cases this is the most massive \subhalo. \ Other \subhalos\ are referred to as satellite \subhalos.

\subsubsection{Halo occupation distribution}

%% \begin{figure}
%%   \centering
%%   \includegraphics[width=\linewidth]{Figures/occupancy.png}
%%   \caption{A histogram of the occupancy of host \halos\ in \Name\ (i.e.~the number of \subhalos\ that they
%%     contain). This plot contains all host \halos\ and \subhalos\ with $\Np\geq 20$ in all snapshots.  \peter{Do we need this Figure as the next one is better?}}
%%   \label{fig:subocc}
%% \end{figure}

\begin{figure}
  \centering
  \includegraphics[width=\linewidth]{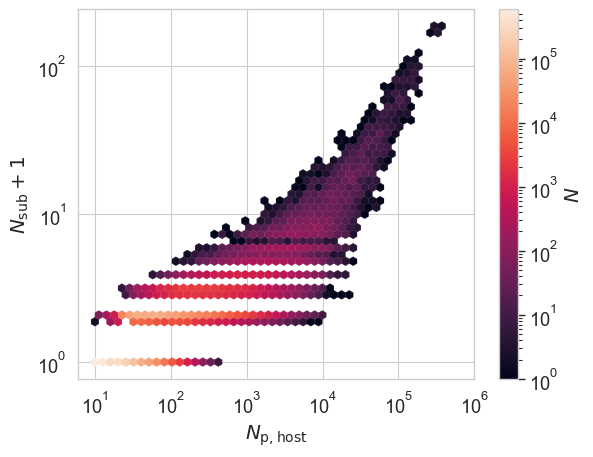}
  \caption{The occupancy as a function of host halo mass in \Name\ for all sub\halos\ at all snapshots. The cluster of points at the tip of the distribution is located in a single halo at multiple sequential timesteps.}
  \label{fig:suboccvsmass}
\end{figure}

\Fig{suboccvsmass} shows the relationship between the number of \subhalos, $N_\mathrm{sub}$, and the particle number in the host halo.  By far the most common values of $N_\mathrm{sub}$ are 0 and 1: the former correspond to small \halos\ for which substructure cannot be resolved, and the latter to \halos\ that have a single density peak.  However, we also see occupancy numbers, of decreasing frequency, as high as $N_\mathrm{sub}=140$.\footnote{Note that the clump of points with $N_\mathrm{sub}>100$ in \Fig{suboccvsmass} all correspond to \halos\ at different snapshots along the main branch of a single graph.}  Of course, the number of \subhalos\ is primarily limited by the resolution of the simulation, i.e.~the number of particles in each halo.

\subsubsection{\Subhalo\ mass fractions}

\begin{figure}
  \centering
  \includegraphics[width=\linewidth]{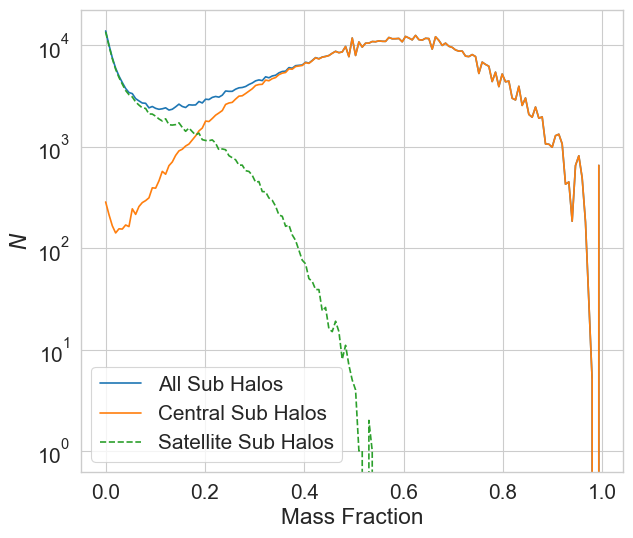}
  \caption{Histograms of \subhalo\ mass fractions, the fraction of a host halo's mass held within the central \subhalo\ and all satellite \subhalos.  The spikiness of the curves, most evident for the central sub-\halos, comes from the discreteness of the particle number in low-mass \halos.}
  \label{fig:submfrac}
\end{figure}

\begin{figure}
  \centering
  \includegraphics[width=\linewidth]{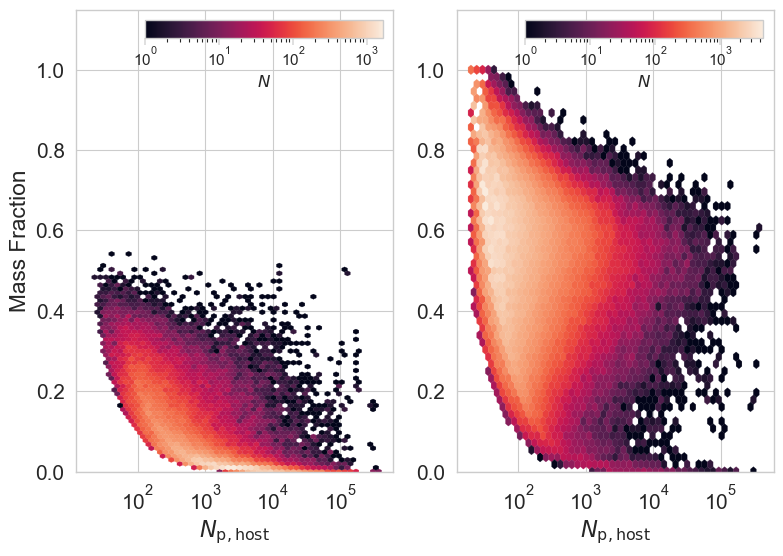}
  \caption{The mass fraction from Fig.\ref{fig:submfrac} plotted as a function of host halo mass. The left panel and right panel show the mass fraction of satellite \subhalos\ and central \subhalos\ respectively.}
  \label{fig:mfracvsmass}
\end{figure}

Figures~\ref{fig:submfrac} and \ref{fig:mfracvsmass} show histograms of the contribution to the mass of a halo from
substructures, and a breakdown by halo mass, respectively.  Note that these plots show only those \halos\ that possess
\subhalos: this is about half of the total.  The requirement that a \subhalo\ has a minimum mass of 20
particles leads to the lower locus seen in the plots and biases the mass fractions upwards in the lowest mass \halos.
Overall the mass fraction of the central \subhalo\ is about 50 per cent but with a large dispersion.  That leaves little
room for satellite \subhalos\ that mostly have very small mass fractions.

The right-hand panel in \Fig{mfracvsmass} shows a branch of very low mass fractions for the central halo -- that is because there are a small number of merging systems for which the most massive \subhalo\ is not located at the centre of the FOF group.  Were we to choose to define the 'main' \subhalo\ as the most massive one then this low mass fraction branch would disappear, but we prefer not to do that, for the reason discussed in the next subsection.

\subsubsection{\Subhalo\ centrality}

\begin{figure}
  \centering
  \includegraphics[width=\linewidth]{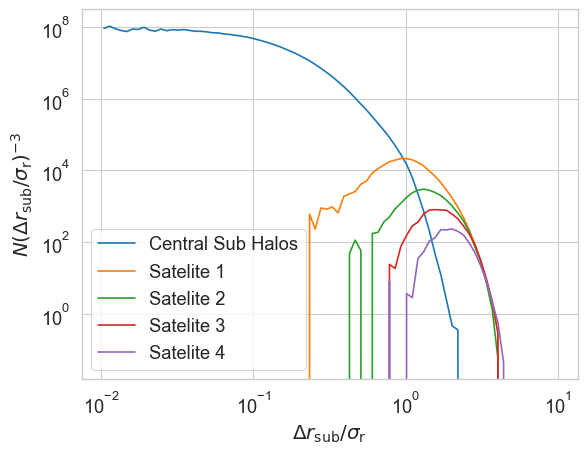}
  \caption{Space density of the centrality of \subhalos\ for all host \halos\ and \subhalos\ with $\Np\geq 10$ in all snapshots. The centrality is the separation of a \subhalo's centre of mass from the centre of mass of its enclosing host halo, $\Delta r_\mathrm{sub}$, divided by the root-mean-square size of the host halo, $\sigma_r$. Plotted here are the central \subhalo\ and then the next 4 subsequent satellite \halos\ in increasing radius from the centre of the host halo.}
  \label{fig:centrality}
\end{figure}

\begin{figure}
  \centering
  \includegraphics[width=\linewidth]{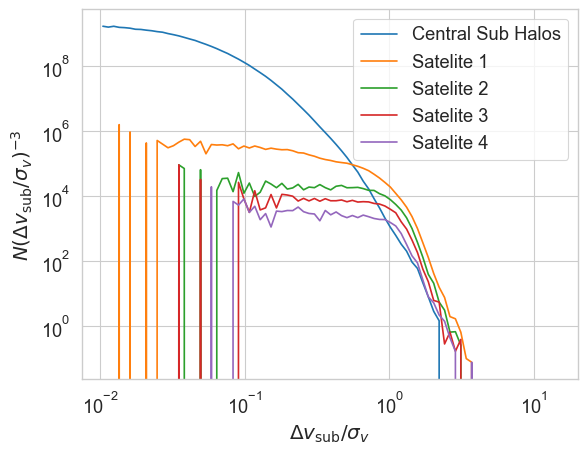}
  \caption{Velocity-space density of the relative speed of central and satellite \subhalos\ for all host \halos\ and \subhalos\ with $\Np\geq 10$ in all snapshots. The relative speed is the magnitude of the difference in velocity of a \subhalo\ from that of the host halo, $\Delta v_\mathrm{sub}$, divided by the velocity dispersion of the host halo, $\sigma_v$.  Plotted here are the central \subhalo\ and then the next 4 subsequent satellite \halos\ in increasing radius from the centre of the host halo.}
  \label{fig:ventrality}
\end{figure}

One of the advantages of defining {\bf all} \subhalos, including the central \subhalo\ as being distinct from the enclosing host halo is that we can determine whether
or not the two have common dynamics, or whether the central \subhalo\ is displaced or is moving with respect to the host: this is shown in Figures~\ref{fig:centrality} and \ref{fig:ventrality}, respectively.\footnote{The scaling of number count with inverse volume in either configuration- or velocity-space in Figures~\ref{fig:centrality} and \ref{fig:ventrality} gives the density of \halos.}

We can see that central \halos\ completely dominate at small radii and remain the most populous type of \subhalo\ all the way out to \Rhalo, where \Rhalo\ is the root-mean-square size of the host halo.  Most likely satellite \subhalos\ are disrupted at smaller radii, although there may be some that survive but go undetected by our halo-identification algorithm.  The median offset of the central \subhalo\ from the centroid of the host halo is about 0.27\,\Rhalo.

In velocity space there is no such restriction and the density of satellite \subhalos\ remains constant, albeit much lower than that of the main \subhalo, all the way in to relative velocities of zero.  The median speed of the central \subhalo\ relative to that of the host halo is about 0.12\,$\sigma_v$.

\section{Conclusions and discussion}
\label{sec:conc}

In this paper we advocate the use of merger graphs rather than merger trees in order to follow the development of structure in the Universe.  Our method leads to two significant advantages compared to previous approaches:
\begin{shortitem}
\item the elimination of catastrophic failures in merger histories whereby massive \halos\ either disappear or are created without progenitors.
\item the reduction of mass fluctuations during the growth of structure;
\end{shortitem}

We identify \halos\ by a two-step process: first using friends-of-friends (FOF) to locate over-dense regions in real (configuration) space; then by running FOF a second time, in phase space, with a decreasing velocity-space linking length, until a candidate halo either has negative total energy or drops below the 10-particle limit.  Although this method does an excellent job at distinguishing interacting but unbound \halos\ that will later separate, it is not perfect.  Therefore, in addition, a small fraction of \halos\ with positive energy are retained if they are descendants of (have 10 or more particles in common with) real \halos\ in the previous snapshot.

To aid comparison with previous work \citep[][SMT13]{Srisawat13}, we also generate merger trees by splitting \halos\ with multiple descendants into 2 or more pieces.  We do not require that the pieces have negative energy, but we do require that they occupy distinct regions of phase-space as an indicator that they are genuine structures.  Our main results are:
\begin{shortitem}
\item The distribution of main branch length (length of time that a halo exists for) is much more peaked than in other methods from SMT13 and in particular shows many fewer short lengths which are indicative of catastrophic failures (\halos\ being created or disappearing out of nothing).
\item The number of progenitors is similar to the SMT13 algorithms.
\item The mass growth along the main branch is slightly more peaked and biased to positive values, but the distribution of mass fluctuations is slightly broader: this is similar to the result found by \citet{Avila14} who showed that the difference arises from the difference between FOF \halos\ and the use of spherical overdensity as in SMT13.
\end{shortitem}

The main purpose of this paper is to look at the properties of merger graphs, with the following conclusions:
\begin{shortitem}
\item As for our derived merger trees, the advantages of increased branch lengths compared to previous methods from SMT13 are retained.
\item In addition the graphs show smoother mass growth than any tree algorithm.
\item The most common number of descendants for a halo is 1, as one might expect, but can rise as high as 43.
\end{shortitem}

\begin{figure}
  \centering
  \includegraphics[width=\linewidth]{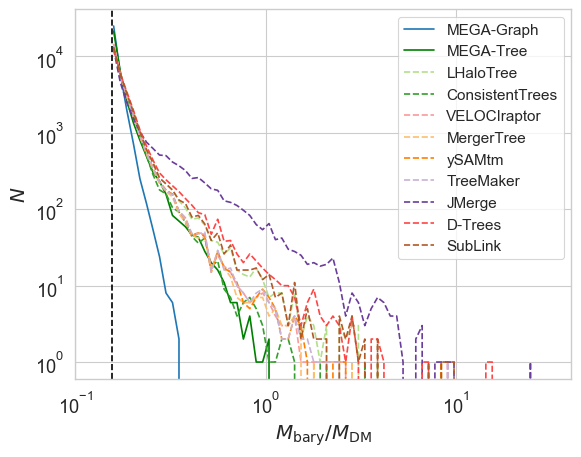}
  \caption{Histograms showing the baryon fraction within current-day \halos\ for merger trees and merger graphs.  The vertical, dashed line shows the mean baryon fraction of the Universe.}
  \label{fig:baryonfraction}
\end{figure}

When modelling star formation in galactic \halos, it is often assumed that the baryon content of \halos\ can never decrease: this is to preserve metallicity \citep[see, e.g.~][]{Henriques20}.  That means that the baryon content of a halo is determined by the ratio of the maximum mass along its past history to the current mass.  \Fig{baryonfraction} shows histograms for the baryon fraction in current-day \halos\ derived from 2 different trees and \Graph, where the baryon mass is set equal to the mean baryon density parameter of the Universe (set to 0.0155 here) times this maximum mass.  In each case, we have strong peaks at the desired universal value, but with a tail of mass ratios extending towards the right: for the best of the SMT13 algorithms, this can even give baryon masses that exceed the total mass of the halo!  The \Tree\ algorithm performs better than any of those from SMT13, but clearly \Graph\ greatly improves on any of the trees.

The \halos\ that we identify with a linking length of 0.2 times the mean separation we call host \halos: we then generate high-density '\subhalos' through the same procedure as for host \halos\ but using half the spatial linking length.  This guarantees nesting of \subhalos\ within main \halos.  The lowest mass host \halos\ have either a single \subhalo\ or none at all, with the number of \subhalos\ increasing with host halo mass up to, for the SMT13 simulation, over 100.

We divide the \subhalos\ up into central \subhalos\ (those closest to the centre of each host halo) and satellite \subhalos.  Typically a fraction of about 0.5-0.6 of the mass is contained in the central \subhalo.  As might be expected, the central \subhalo\ is generally located towards the centre of the enclosing main halo, with a median separation between their centres of about 0.27\,\Rhalo, where \Rhalo\ is the root-mean-square size of the host halo.  Similarly, most central \subhalos\ are moving at roughly the same velocity as their hosts.  In a small fraction of cases, however, corresponding to merging systems, there are substantial offsets between the positions and/or velocities of central \subhalos\ relative to their hosts.

In the context of galaxy formation then we associate \halos\ with the potential wells (extended galactic \halos, groups and clusters) within which baryons will accumulate.  The galaxies themselves will form within the higher-density \subhalos.  The central \subhalo\ is usually, but not always, located at the centre of the host halo and will be the natural site of accretion for cooling coronal gas: where there is a large offset then that is an indicator of a merger and accretion could be suspended.  

Thus we anticipate that our merger graphs will provide a better backbone for galaxy formation models than existing merger trees in three respects:
\begin{shortitem}
  \item They eliminate critical failures in the halo growth history: massive \halos\ with either no progenitors or no descendants.
\item They provide smoother mass growth for the graph as compared to individual trees.
\item They distinguish between the central \subhalo\ and the host halo (unlike other methods) and thus provide a better treatment of mergers.
\end{shortitem}
For existing semi-analytic models that rely on merger trees rather than merger graphs then the \Tree\ algorithm greatly improves with respect to other trees on the first of these properties and also satisfies the third. (We note, however, that, unlike the graphs, trees cannot be constructed on-the-fly as \Tree\ requires that we start with the final snapshot and work back in time.)

We intend to investigate the use of \Graph\ and \Tree\ for galaxy formation modelling in a future paper.

%%%%%%%%%%%%%%% ACKNOWLEDGEMENTS %%%%%%%%%%%%%%%%%%
\section*{Acknowledgements}

The authors contributed in the following way to this paper: PAT originated and supervised the project; WJR did the vast majority of the coding and produced the figures; PAT and WJR together drafted the paper; CS worked on an early version of the project, chipped in with useful suggestions from time to time, and helped to improve the paper.

WJR was jointly supported by a Royal Astronomical Society summer student bursary and by a Junior Research Scholarship from the University of Sussex for a substantial part of this project.  PAT %(ORCID 0000-0001-6888-6483)
acknowledges support from the Science and Technology Facilities Council (grant number ST/P000525/1). CS acknowledges support by a Grant of Excellence from the Icelandic Research Fund (grant number 173929-051).

This project made extensive use of data from the Sussing Merger Trees project \citep{Srisawat13}, which is available upon request from the authors: we would like to thank everyone who contributed to that project.

We have benefited greatly from the publicly available programming language {\tt python}, including the {\tt numpy}, {\tt matplotlib}, {\tt scipy} and {\tt h5py} packages. With thanks to \cite{Thompson2014}\footnote{\tt https://bitbucket.org/rthompson/pygadgetreader/}, used to read {\tt GADGET} binary files.

This work used the DiRAC@Durham facility managed by the Institute for Computational Cosmology on behalf of the STFC DiRAC HPC Facility (www.dirac.ac.uk). The equipment was funded by BEIS capital funding via STFC capital grants ST/K00042X/1, ST/P002293/1, ST/R002371/1 and ST/S002502/1, Durham University and STFC operations grant ST/R000832/1. DiRAC is part of the National e-Infrastructure.

In addition, much of the early analysis was undertaken on the {\sc Apollo} cluster at Sussex University. 
  
\bibliographystyle{mnras}
\bibliography{mergergraph}

%%%%%%%%%%%%%%%%% APPENDICES %%%%%%%%%%%%%%%%%%%%%
%
\appendix

\section{Omission of 10 particle haloes}
\label{sec:10part}

\begin{figure}
  \centering
  \includegraphics[width=\linewidth]{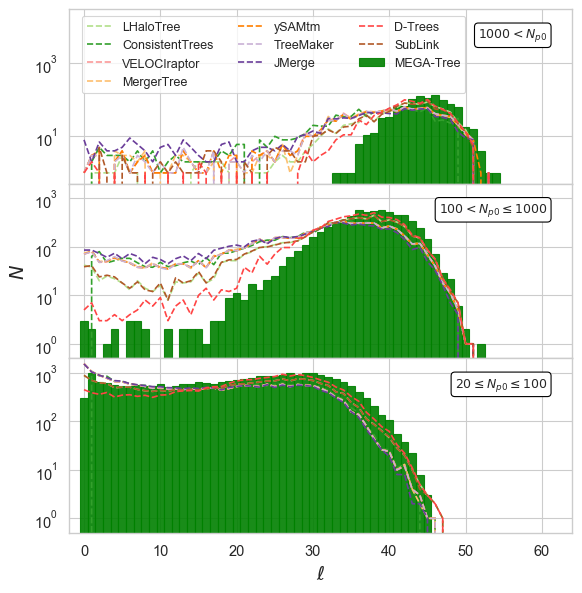}
  \caption{The main branch length of all trees rooted at the present day ($z=0$) omitting $N_{p}<20$ \halos.  $\ell$ is the number of snapshots that a tree's main branch can be traced until it disappears. The upper, middle and lower panels represent mass bins of $\Np > 1000$, $1000\geq\Np>100$ and $100\geq\Np\geq20$ respectively.}
  \label{fig:treemainbranch20}
\end{figure}

\begin{figure}
  \centering
  \includegraphics[width=\linewidth]{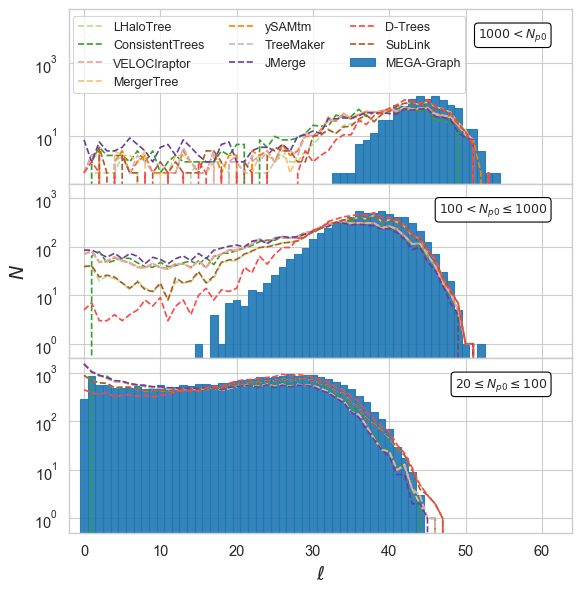}
  \caption{The main branch length of all graphs rooted at the present day ($z=0$) omitting $N_{p}<20$ \halos.  $\ell$ is the length of a main branch, and the lower, middle, and upper panels represent mass bins of $20\leq \Np \leq 100$, $100< \Np \leq 1000$ and $1000\leq \Np$ respectively. The \Name\-Graph main branches are defined such that they are rooted at generations of the graph where the most massive halo has a mass bounded by the upper and lower limit of the respective mass bin.}
  \label{fig:graphmainbranch20}
\end{figure}

We wish to emphasise that the strength of both \Name-Tree and \Name-Graph is not due to the mass limit of $N_{p}>10$. Instead their strength comes from the encoding of information from an arbitrary number descendants, inherent in the construction of a graph, and subsequent teasing apart into trees. To this end, we have regenerated our results using only halos with $\Np\geq20$, for more direct comparison with the results from SMT13.   We present the main branch lengths in \Fig{treemainbranch20} and \Fig{graphmainbranch20} for the trees and graphs respectively.

Strikingly, in both the graphs and trees we retain the behaviour observed in \Fig{treemainbranch} and \Fig{graphmainbranch} in the middle and upper panel. In the case of \Fig{graphmainbranch20} this definitively shows the robustness of the merger graph approach we advocate. However, for \Name-Tree the retention of the improved behaviour in these mass bins may come as a surprise. Once again, this improvement can be attributed to the inclusion of descendent information in merger graphs which is then inherited by the trees via the splitting of \halos\ with multiple descendants.

In the lower mass bin, however, the graph approach does not provide a significant improvement but gives results similar to the SMT13 algorithms, with the exception of D-Trees which tracks fluctuations at the low mass limit better than others allowing snapshots to be skipped when looking for progenitors. Omitting these $N_{p}<20$ \halos\ removes our mechanism to compensate for downwards mass fluctuations in low-mass \halos.

To have well behaved main branch lengths in the low mass regime it is therefore imperative to have some mechanism to track fluctuations around the mass limit of a halo definition, be it snapshot skipping as utilised in D-Trees or tracking \halos\ below the halo mass definition as employed here.

\section{Flowcharts of the algorithm}
\label{sec:flows}

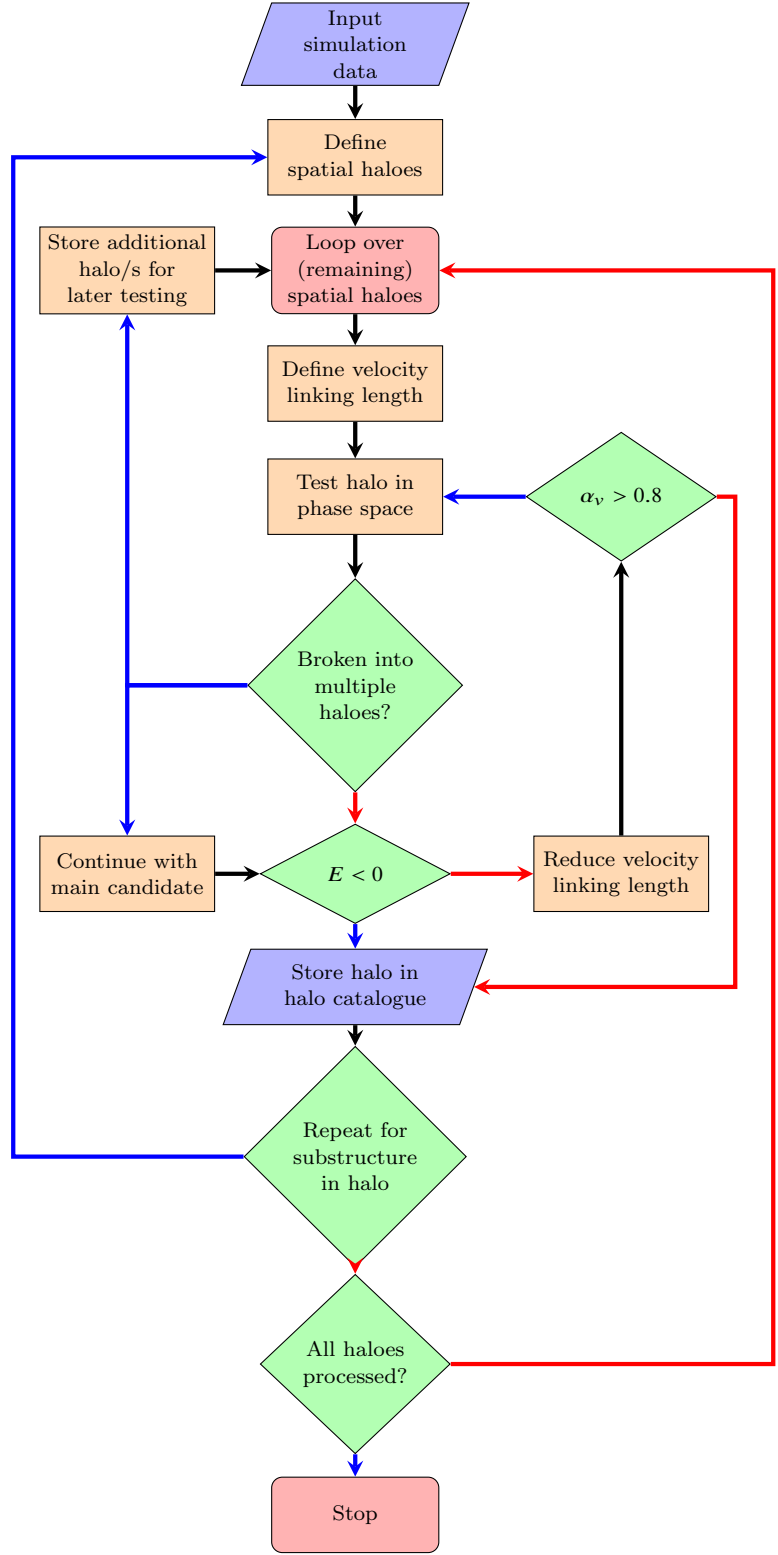
\begin{figure*}
\centering
\begin{tikzpicture}[node distance=1.5cm]

\node (in) [io, align=center] {Input \\simulation data};
\node (spatial) [process, below of=in] {Define \\spatial \halos};
\node (loop) [startstop, below of=spatial] {Loop over \\ (remaining) spatial \halos};
\node (vellinkl) [process, below of=loop] {Define velocity linking length};
\node (storesplithalo) [process, left of=loop, xshift=-1.5cm] {Store additional halo/s for later testing};
\node (phasetest) [process, below of=vellinkl] {Test halo in phase space};
\node (split) [decision, below of=phasetest, align=center, yshift=-1.0cm] {Broken into \\multiple \\\halos?};
\node (limitvelinkl) [decision, right of=phasetest, xshift=2.0cm] {$\alpha_{v}>0.8$};
\node (E2) [decision, below of=split, yshift=-1.0cm] {$E<0$}; 
\node (cont) [process, left of=E2, xshift=-1.5cm] {Continue with main candidate};
\node (redefvlinkl) [process, right of=E2, xshift=2.0cm] {Reduce velocity linking length};
\node (out) [io, below of=E2] {Store halo in halo catalogue};
\node (subhalo) [decision, below of=out, yshift=-0.75cm, align=center] {\small Repeat for \\ substructure \\ in halo};
\node (next) [decision, below of=subhalo, yshift=-1.25cm, align=center] {All \halos\  \\processed?};
\node (stop) [startstop, below of=next, yshift=-0.5cm] {Stop};

\draw [arrow] (in) -- (spatial);
\draw [arrow] (spatial) -- (loop);
\draw [arrow] (loop) -- (vellinkl);
\draw [arrow] (storesplithalo.east) -- (loop.west);
\draw [arrow] (vellinkl) -- (phasetest);
\draw [arrow] (phasetest) -- (split);
\draw [arrow, red] (split) -- (E2);
\draw [arrow, blue] (E2.south) -- (out.north);
\draw [arrow, red] (E2.east) -- (redefvlinkl.west);
\draw [arrow, blue] (split.west) -| (storesplithalo.south);
\draw [arrow, blue] (split.west) -| (cont.north);
\draw [arrow] (cont.east) -- (E2.west);
\draw [arrow] (redefvlinkl) -- (limitvelinkl.south);
\draw [arrow, blue] (limitvelinkl) -- (phasetest.east);
\draw [arrow] (out.south) -- (subhalo.north);
\draw [arrow, red] (subhalo.south) -- (next.north);
\draw [arrow, red] (limitvelinkl.east) --(5.0cm, -6.0cm) |- (out.east);
\draw [arrow, blue] (subhalo.west) --(-4.5cm, -14.75cm) |- (spatial.west);
\draw [arrow, red] (next.east) --(5.5cm, -17.5cm) |- (loop.east);
\draw [arrow, blue] (next.south) --(stop.north);
\end{tikzpicture}
\caption{A flowchart of the halo finder algorithm. This process is performed on each snapshot independently, producing the halo catalogue which is later pruned at the merger graph construction phase: eliminating \halos\ with $E>0$ and $N<20$ that do not have any progenitors. In the flow chart, blue trapeziums represent I/O, red rounded boxes represent loops, orange boxes represent processes and green diamonds represent logic. Blue arrows represent True/Yes, while red arrows represent False/No.}
\label{fig:haloflow}
\end{figure*}

\begin{figure*}
\centering
\begin{tikzpicture}[node distance=1.5cm]

\node (ink) [io] {Input snapshot $k$ and $k+1$ \halos};
\node (loop) [startstop, below of=ink, yshift=-0.25cm] {Loop over \\(remaining) snapshot $k$ \halos};
\node (descs) [process, below of=loop, yshift=-0.25cm] {Identify \\descendants};
\node (multidesc) [decision, below of=descs, yshift=-0.5cm] {$N_{\mathrm{desc}}>1$};
\node (nosplit) [process, right of=multidesc, xshift=1.5cm] {Halo does not need splitting};
\node (identmain) [process, below of=multidesc, yshift=-0.75cm] {Identify main splitting candidate from descendant contribution};
\node (splitloop) [startstop, below of=identmain, yshift=-0.25cm] {Loop over splitting candidates};
\node (main) [decision, below of=splitloop, yshift=-0.25cm] {Main?};
\node (boundsplit) [decision, below of=main, yshift=-0.25cm] {$E<0$};
\node (hold) [process, right of=boundsplit, xshift=2.25cm] {Hold candidate \\ to be \\ written out};
\node (ovsplit) [decision, below of=boundsplit, yshift=-0.5cm] {Overlap?};
\node (finished) [decision, right of=splitloop, xshift=5cm, align=center] {Finished \\ splitting?};
\node (recombine) [process, above of=finished, yshift=0.75cm] {Calculate \\split-halo masses from graph halo mass};
\node (splitout) [io, right of=nosplit, xshift=2.0cm] {Write out post-splitting \halos};
\node (holdnot) [process, below of=ovsplit, yshift=-0.25cm] {Discard candidate};
\node (next) [decision, right of=loop, align=center, xshift=5.0cm] {All \halos\  \\processed?};
\node (stop) [startstop, right of=next, xshift=1.5cm] {Stop};

\draw [arrow] (ink) -- (loop);
\draw [arrow] (loop) -- (descs);
\draw [arrow] (descs.south) -- (multidesc.north);
\draw [arrow] (hold.east) -| (finished.south);
\draw [arrow] (recombine) -- (splitout);
\draw [arrow] (nosplit) |- (splitout.west);
\draw [arrow] (splitout) -- (next);
\draw [arrow, red] (next) -- (loop);
\draw [arrow, blue] (next) -- (stop);
\draw [arrow, red] (finished.west) -- (splitloop.east);
\draw [arrow, blue] (finished.north) -- (recombine.south);
\draw [arrow, red] (multidesc.east) -- (nosplit.west);
\draw [arrow, blue] (multidesc.south) -- (identmain.north);
\draw [arrow] (identmain) -- (splitloop);
\draw [arrow] (splitloop) -- (main);
\draw [arrow, red] (main) -- (boundsplit);
\draw [arrow, blue] (main.east) -| (hold.north);
\draw [arrow, blue] (boundsplit.east) -- (hold.west);
\draw [arrow, red] (boundsplit.south) -- (ovsplit.north);
\draw [arrow, red] (ovsplit.east) -| (hold.south);
\draw [arrow, blue] (ovsplit.south) -| (holdnot.north);
\draw [arrow] (holdnot) -| (finished);
\end{tikzpicture}
\caption{A flowchart of the graph splitting algorithm used to build merger trees from merger graphs. This is done on a snapshot by snapshot basis, starting at the penultimate snapshot and working back in time. In the flow chart, blue trapeziums represent I/O, red rounded boxes represent loops, orange boxes represent processes and green diamonds represent logic. Blue arrows represent True/Yes, while red arrows represent False/No.}
\label{fig:splitflow}
\end{figure*}
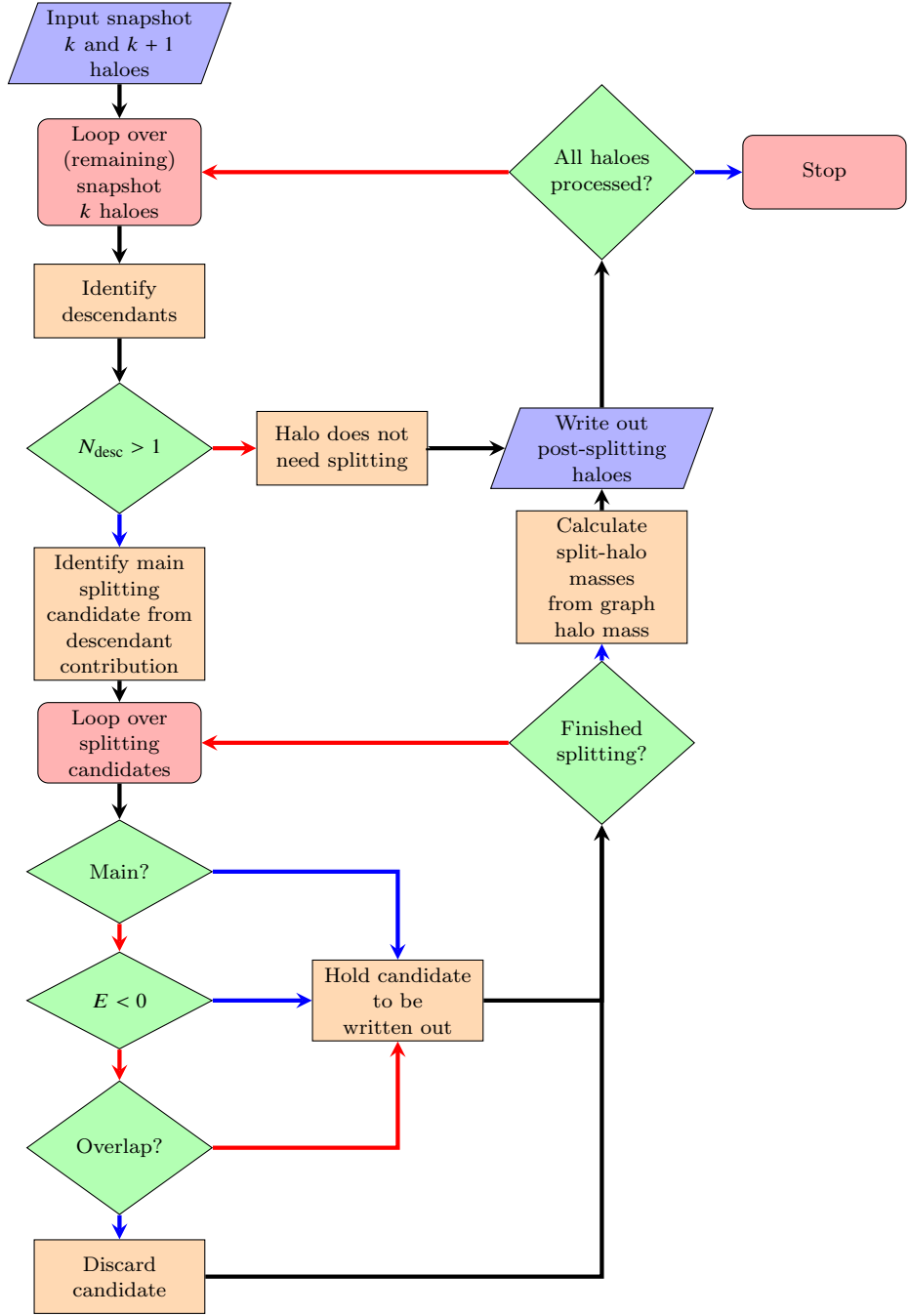

Here we present flowcharts detailing the more complex elements of \Name. We have omitted the method for merger graph construction as this is very straightforward, as described in \Sec{method:graphs}. The halo finder and tree splitting algorithms are presented in \Fig{haloflow} and \Fig{splitflow} respectively.

\section*{Supporting Information}
The merger tree software described in the paper is available at {\tt https://github.com/wjr21/mega}.

%%%%%%%%%%%%%%%%%%%%%%%%%%%%%%%%%%%%%%%%%%%%%%%%%%
% Don't change these lines
\bsp	% typesetting comment
\label{lastpage}
\end{document}